\documentclass[sigconf]{acmart}

\usepackage{booktabs} 
\usepackage{xcite}
\usepackage{xr}
\usepackage{graphicx}
\usepackage{caption}
\usepackage{subcaption}
\usepackage{multirow}
\usepackage{bigstrut}
\usepackage{enumitem}
\usepackage[skip=0pt]{caption}
\usepackage[ruled,linesnumbered]{algorithm2e}
\usepackage{bm}
\usepackage{tabularx,ragged2e}
\usepackage{xcolor}
\usepackage{verbatim}

\usepackage{amsmath}

\externalcitedocument{introduction}
\externalcitedocument{methods}
\externalcitedocument{evaluation}
\renewcommand\footnotetextcopyrightpermission[1]{} 
\makeatletter
\def\subsubsection{\@startsection{subsubsection}{3}%
  \z@{.3\linespacing\@plus.7\linespacing}{.1\linespacing}%
  {\normalfont\itshape}}
\makeatother

\copyrightyear{2019} 
\acmYear{2019} 
\setcopyright{acmcopyright}
\acmConference[KDD '19]{The 25th ACM SIGKDD Conference on Knowledge Discovery and Data Mining}{August 4--8, 2019}{Anchorage, AK, USA}
\acmBooktitle{The 25th ACM SIGKDD Conference on Knowledge Discovery and Data Mining (KDD '19), August 4--8, 2019, Anchorage, AK, USA}
\acmPrice{15.00}
\acmDOI{10.1145/3292500.3330984}
\acmISBN{978-1-4503-6201-6/19/08}

\begin{document}
\title{Hierarchical Gating Networks for Sequential Recommendation}

\author{Chen Ma}
\affiliation{\institution{School of Computer Science\\McGill University}}
\email{chen.ma2@mail.mcgill.ca}

\author{Peng Kang}
\affiliation{\institution{Department of Computer Science\\Northwestern University}}
\email{pengkang2022@u.northwestern.edu}

\author{Xue Liu}
\affiliation{\institution{School of Computer Science\\McGill University}}
\email{xueliu@cs.mcgill.ca}

\begin{abstract}
The chronological order of user-item interactions is a key feature in many recommender systems, where the items that users will interact may largely depend on those items that users just accessed recently. However, with the tremendous increase of users and items, sequential recommender systems still face several challenging problems: (1) the hardness of modeling the long-term user interests from sparse implicit feedback; (2) the difficulty of capturing the short-term user interests given several items the user just accessed. To cope with these challenges, we propose a hierarchical gating network (HGN), integrated with the Bayesian Personalized Ranking (BPR) to capture both the long-term and short-term user interests. Our HGN consists of a feature gating module, an instance gating module, and an item-item product module. In particular, our feature gating and instance gating modules select what item features can be passed to the downstream layers from the feature and instance levels, respectively. Our item-item product module explicitly captures the item relations between the items that users accessed in the past and those items users will access in the future. We extensively evaluate our model with several state-of-the-art methods and different validation metrics on five real-world datasets. The experimental results demonstrate the effectiveness of our model on Top-N sequential recommendation.
\end{abstract}



\maketitle

\section{Introduction}
As the Internet service and mobile device usages keep growing, Internet users can easily access a large number of online products and services. Although this growth provides users with more available choices, it is also difficult for users to pick up one of the most favorite items out of plenty of candidates. To reduce information overload and satisfy the diverse needs of users, personalized recommender systems come into being and play more and more important roles in modern society. These systems can provide personalized experiences, serve huge service demands, and bring significant benefits to at least two parties: (1) help users easily discover products that they are interested in; (2) create opportunities for product providers to increase the revenue.

In all kinds of Internet services, users access the products or items in a chronological order, where the items a user will interact may be closely relevant to those items she just accessed. This property facilitates a non-trivial recommendation task---sequential recommendation, which treats the user behavior history as an action sequence ordered by the operating timestamp. This task is challenging to address due to one major reason: the difficulty of inferring users' short-term interests and intentions. Indeed, both the long-term and short-term interests of users together determine the users' actions on items. With the large accumulated data, the long-term user interests can be effectively modeled. However, within a short-term context, how to take advantage of the sequential dynamics for predicting user actions in the near future is non-trivial.

To capture the sequential dynamics in the user action history, effective models are proposed to learn the short-term user preference in the sequential user interactions, such as Markov Chains (MCs), convolutional neural networks (CNNs), and recurrent neural networks (RNNs). MC-based methods \cite{DBLP:conf/www/RendleFS10,DBLP:conf/icdm/HeM16} apply a $ K $-order Markov chain to make recommendations based on the $ K $ previous actions. CNN-based methods \cite{DBLP:conf/wsdm/TangW18} utilize convolutional filters and sliding window strategies to capture the short-term contexts for future prediction. RNN-based methods \cite{DBLP:journals/corr/HidasiKBT15,DBLP:conf/recsys/QuadranaKHC17,DBLP:conf/cikm/HidasiK18} adopt gated recurrent (GRU) or long short-term memory (LSTM) units to learn the user-item sequence, where the short-term user interests are captured by the hidden states of RNNs.

Although existing methods have proposed effective models and achieved satisfactory results, we argue that there are still several factors to be considered for enhancing the performance. First, previous studies \cite{DBLP:conf/wsdm/TangW18,DBLP:journals/corr/HidasiKBT15,DBLP:conf/recsys/QuadranaKHC17,DBLP:conf/cikm/HidasiK18} learn the user action sequence by CNN or RNN structures, which does not consider the specific parts of features of different items. Neglecting the representative features may fail to capture the true user interests in a short context. Second, these CNN or RNN based methods also do not discriminate the item importance based on users' preferences. Equally treating those informative items along with other items may lead to the incomplete understanding of user intentions. Third, it is also important to note that the relations between items are neglected in previous works \cite{DBLP:conf/icdm/KangM18,DBLP:conf/cikm/HidasiK18,DBLP:conf/wsdm/TangW18}. It is very likely that closely related items may be interacted by users one after the other. As such, explicitly capturing the item-item relations will largely benefit predicting subsequent items users will interact.

To address the problems mentioned above, we propose a novel recommendation model, hierarchical gating network (HGN), for the sequential recommendation task without using complex recurrent or convolutional neural networks. HGN consists of a feature gating module, an instance gating module, and an item-item product module, integrated with the matrix factorization model and optimized by the Bayesian Personalized Ranking (BPR) objective. In particular, the feature gating module allows the adaptive selections of attractive latent features of items based on the user preference, where the selected user-specific features will be passed to the instance gating module. At the instance gating module, important items that reflect the short-term user interests will be distinguished and selected for future item prediction. Thus, the feature gating and instance gating modules form a hierarchical gating network to control what features or items can be passed to the downstream layers. On the other hand, item-item relations provide important auxiliary information to predict users' sequential behaviors, since closely related items may be interacted by users one after the other. Thus, we apply an item-item product module to explicitly capture the relations between the items users have interacted and those items user will interact in the future. We extensively evaluate our model with many state-of-the-art methods and different validation metrics on five real-world datasets. The experimental results not only demonstrate the improvements of our model over other baselines but also show the effectiveness of the gating and item-item product modules.

To summarize, the major contributions of this paper are listed as follows:
\begin{itemize}[leftmargin=*]
\item To infer the user interests in a short-term context, we propose a hierarchical gating network to control what item latent features and which relevant item can be passed to the downstream layers. Our hierarchical gating network achieves better performance compared with complex recurrent or convolutional neural networks yet with fewer parameters and faster training speed.
\item To explicitly capture the item-item relations, we utilize an item-item product module to learn the relationships between the items users have interacted and those items user will interact in the near future.
\item Experiments on five real-world datasets show that the proposed HGN model significantly outperforms the state-of-the-art methods for the sequential recommendation task.
\end{itemize}

\section{Related Work}
In this section, we illustrate related work about the proposed model: personalized recommendation with user implicit feedback and the sequential recommendation.

\subsection{Recommendation with Implicit Feedback}
In many real-world recommendation scenarios, user implicit data \cite{DBLP:conf/kdd/WangWY15,DBLP:conf/cikm/TranLL018}, e.g., browsing or clicking history, is more ubiquitous and common than the explicit feedback \cite{DBLP:conf/www/SarwarKKR01,DBLP:conf/icml/SalakhutdinovMH07}, such as user ratings. The implicit feedback only provides positive samples, which is also called one-class collaborative filtering (OCCF) \cite{DBLP:conf/icdm/PanZCLLSY08}. Effective methods are proposed to tackle the OCCF problem. Early works either apply a uniform weighting scheme to treat all missing data as negative samples \cite{DBLP:conf/icdm/HuKV08}, or sample negative instances from missing data to learn the pair-wise user preference between positive and negative samples \cite{DBLP:conf/uai/RendleFGS09}. Recently, several works \cite{DBLP:conf/sigir/HeZKC16,DBLP:conf/www/LiangCMB16} are proposed to weigh the missing data. In \cite{DBLP:conf/www/TranLLK19,DBLP:conf/www/HsiehYCLBE17}, metric learning is applied to compute the distance between users and items. With the ability to represent non-linear and complex data, (deep) neural networks have been utilized in the domain of recommendation and bring more opportunities to reshape the conventional recommendation architectures. In \cite{DBLP:conf/wsdm/WuDZE16,DBLP:conf/cikm/MaZWL18,DBLP:conf/wsdm/MaKWWL19}, (denoising) autoencoders are proposed capture the user-item interaction from user implicit feedback. In \cite{DBLP:conf/www/HeLZNHC17}, He et al. propose a neural network-based collaborative filtering model, where a multi-layer perceptron is utilized to learn the non-linear user-item interactions. In \cite{DBLP:conf/ijcai/XueDZHC17,DBLP:conf/ijcai/GuoTYLH17,DBLP:conf/kdd/LianZZCXS18}, conventional matrix factorization and factorization machine methods are also benefited by the representation ability of deep neural networks.

\subsection{Sequential Recommendation}
Some early sequential recommendation methods rely on item-item transition matrices to capture the sequential patterns in the user interaction sequence. The Markov chain \cite{DBLP:conf/ijcai/ChengYLK13} is a classical option to solve this problem. For example, Rendle et al. \cite{DBLP:conf/www/RendleFS10} propose to factorize personalized Markov chains for capturing long-term preferences and short-term transitions. He et al. \cite{DBLP:conf/icdm/HeM16} combines similarity-based models with high-order Markov chains to make personalized sequential recommendations. In \cite{DBLP:conf/recsys/HeKM17}, the translation-based method is proposed for sequential recommendation. Recently, benefited by the advantages of sequence learning in natural language processing, (deep) neural network based methods are proposed to learn the sequential dynamics. For instance, Tang et al. \cite{DBLP:conf/wsdm/TangW18} propose to apply the convolutional neural network (CNN) on item embedding sequence, where the short-term contexts can be captured by the convolutional operations. In \cite{DBLP:journals/corr/HidasiKBT15,DBLP:conf/recsys/QuadranaKHC17,DBLP:conf/cikm/LiRCRLM17,DBLP:conf/cikm/HidasiK18}, recurrent neural network (RNN), especially gated recurrent unit (GRU), based methods are utilized to model the sequential patterns  for the session-based recommendation \cite{DBLP:journals/corr/HidasiKBT15}, where the hidden states of RNNs reflect the summary of the (sub)sequence. On the other hand, self-attention \cite{DBLP:conf/nips/VaswaniSPUJGKP17} exhibits promising performance in sequence learning and is utilized in sequential recommendation. In \cite{DBLP:conf/icdm/KangM18}, Kang et al. propose to leverage self-attention for adaptively considering interacted items. In \cite{DBLP:conf/wsdm/ChenXZT0QZ18,DBLP:conf/sigir/HuangZDWC18}, memory networks \cite{DBLP:conf/nips/SukhbaatarSWF15} are adopted to memorize the important items that will play a role in predicting future user actions.

However, our hierarchical gating network is different from the above studies. We apply feature-level and instance-level gating modules to adaptively control what item latent features and which relevant item can be passed to the downstream layers. While previous works either do not consider the representative items, or only consider the instance-level importance by the attention model but neglecting the feature-level ones. On the other hand, we adopt an item-item product to explicitly capture the relations between the items users have interacted and those items users will access in the near future, where the explicit modeling of item relations is rarely considered in previous works.

\section{Problem Formulation}
The recommendation task considered in this paper takes sequential implicit feedback as training data. The user preference is presented by a user-item sequence in the chronological order $ \mathcal{S}^{i}=(\mathcal{S}_{1}^{i}, \mathcal{S}_{2}^{i},...,\mathcal{S}_{|\mathcal{S}^{i}|}^{i}) $, where $ \mathcal{S}_{j}^{i} $ is an item index that user $ i $ has interacted with. Given the earlier subsequence $ \mathcal{S}^{i}_{1:t} (t < |\mathcal{S}^{i}|) $ of $ M $ users, the problem is to recommend a list of items from $ N $ items to each user and evaluate whether the items in $ \mathcal{S}^{i}_{t:|\mathcal{S}^{i}|} $ will appear in the recommended list.

Here, following common symbolic notation, upper case bold letters denote matrices, lower case bold letters denote column vectors without any specification, and non-bold letters represent scalars. The major symbols are listed in Table \ref{tab:notations}.

\begin{table}[ht]
\centering
\caption{List of notations.}
\label{tab:notations}
\begin{tabular}{ll}
 \hline
$ M $, $ N $ & the number of users and items \\
$ \mathcal{S}^{i} $ & the item sequence of user $ i $ \\
$ \mathbf{S}_{i,l} $ & the embeddings of the $ l $-th subsequence of user $ i $ \\
$ \mathbf{W}_{g_{*}}, \mathbf{w}_{g_{*}} $ & the learnable parameters in the gating layers \\
$ \mathbf{U} $ & the user embedding matrix \\
$ \mathbf{E} $ & the input item embedding matrix \\
$ \mathbf{Q} $ & the output item embedding matrix \\
$ d $ & the dimension of the embeddings \\
$ \hat{r}_{i, j} $ & the prediction score of user $ i $ on item $ j $ \\
$ \lambda $ & the regularization term \\ \hline
\end{tabular}
\vspace{-0.3cm}
\end{table}

\begin{figure*}[t!]
    \centering
    \begin{subfigure}[t]{0.33\textwidth}
        \centering
        \includegraphics[width=\linewidth]{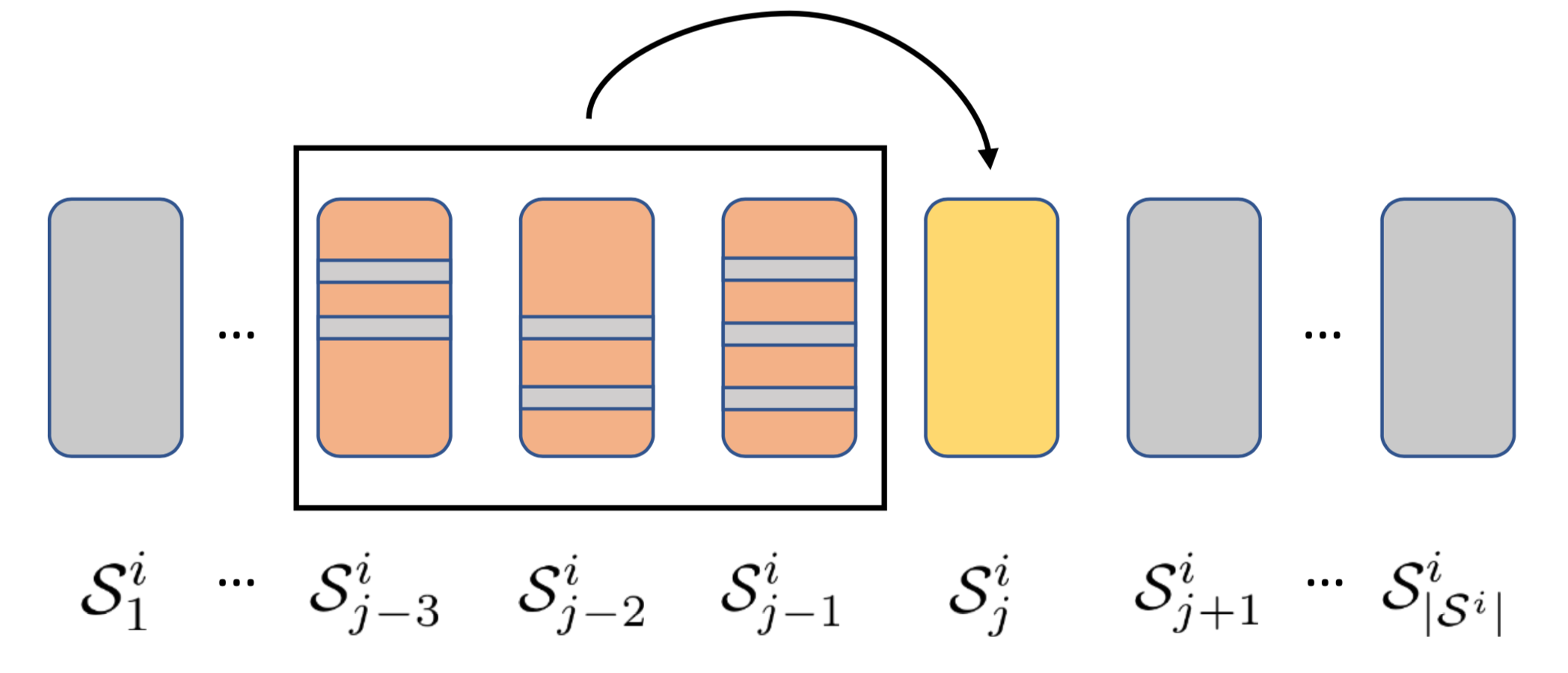}
        \caption{\label{fig:feature_gating}Feature gating}
    \end{subfigure}%
    \begin{subfigure}[t]{0.33\textwidth}
        \centering
        \includegraphics[width=\linewidth]{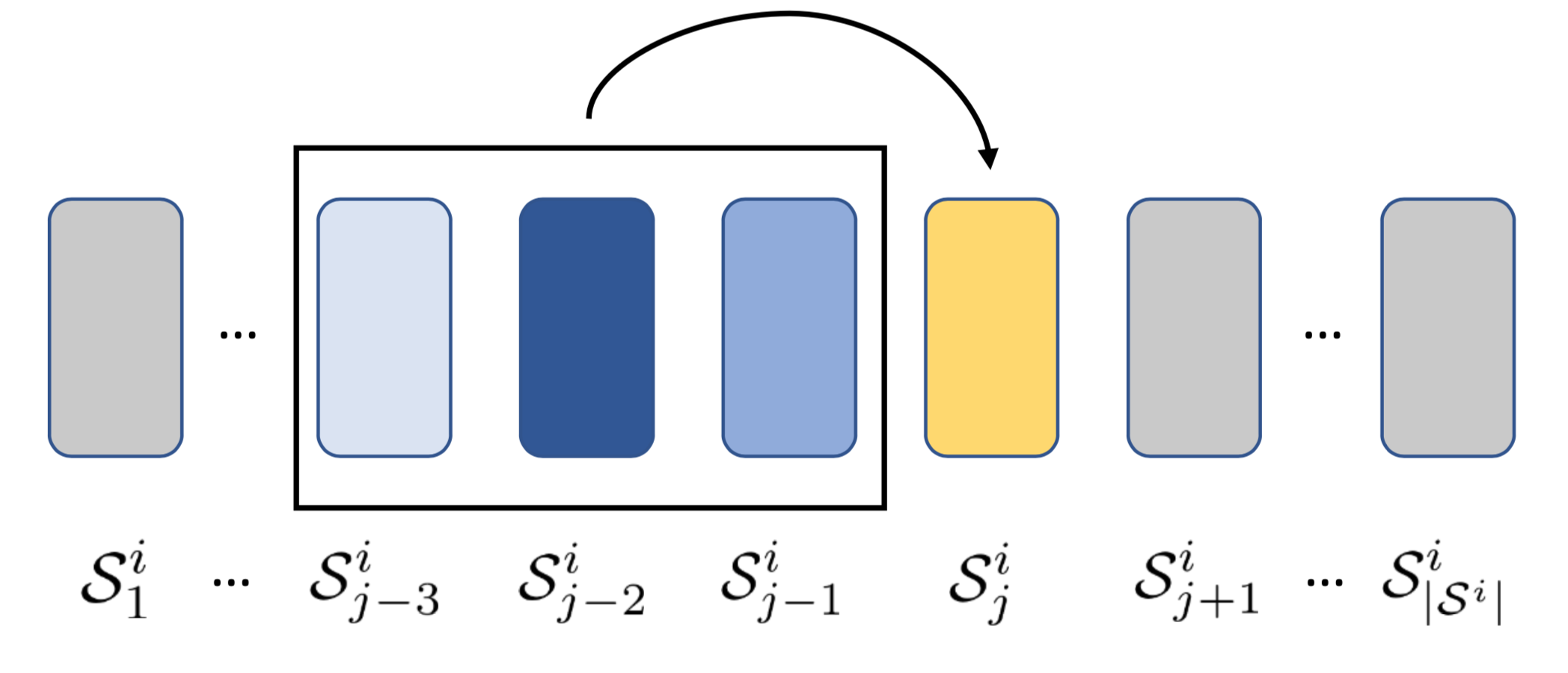}
        \caption{\label{fig:instance_gating}Instance gating}
    \end{subfigure}%
    \begin{subfigure}[t]{0.33\textwidth}
        \centering
        \includegraphics[width=\linewidth]{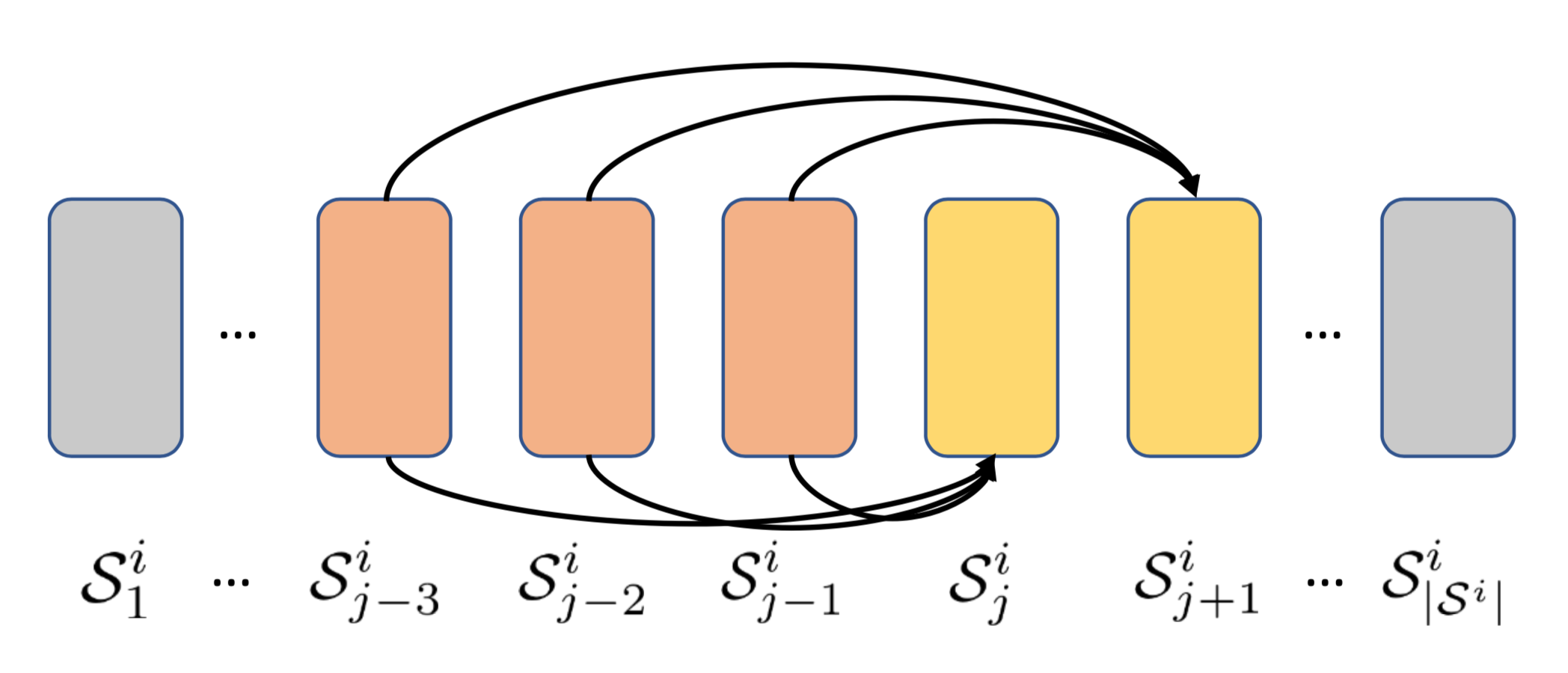}
        \caption{\label{fig:item_product}Item-item product}
    \end{subfigure}
    \Description{An illustrative example}
    \caption{\label{fig:three_modules}An illustrative example of the feature gating, instance gating, and item-item product modules. In Figure \ref{fig:feature_gating}, the gray lines on items denote those latent features are masked off. In Figure \ref{fig:instance_gating}, the darker blue means the item is more important. In Figure \ref{fig:item_product}, the line linked between two items denotes the inner product, which captures the relations between the items users have accessed and the items users will access in the future.}
\end{figure*}

\section{Methodologies}
To model the sequential recommendation task, for each user $ i $, we extract every $ |L| $, i.e. $ L=(\mathcal{S}_{j}^{i}, \mathcal{S}_{j + 1}^{i},...,\mathcal{S}_{j + |L| - 1}^{i}) $, successive items as input and their next $ |T| $ items as the targets to be predicted. The problem can be formulated as: in the user-item interaction sequence $ \mathcal{S}^{i} $, given the $ |L| $ successive items, how likely other $ N $ items will be interacted subsequently.

In the sequential recommendation problem, the prediction of users' preferences on items can be modeled in two perspectives: long-term interests and short-term interests. The long-term user preference modeling has been widely investigated in the conventional Top-N recommendation methods, such as matrix factorization \cite{DBLP:conf/icdm/HuKV08,DBLP:conf/uai/RendleFGS09}. On the other hand, how to capture the short-term user interests from the sequential data is the key point for performance improvement.

For the short-term interest modeling, we argue that there are two kinds of relationships existing between items users have interacted and items users will interact in the future: group-level and instance-level relations. The group-level influence illustrates a phenomenon that several items in $ L $ together have an impact on the items user may interact in the future. For example, if a user has bought a bed frame and a mattress, a pillow is probably a more suitable recommendation than a table. On the other hand, the instance-level influence depicts the strong relation between a single item in $ L $ and a single item in $ T $. For example, if a user bought a mobile phone, she may also need to buy a screen protector or a case. Thus, these two kinds of relations together determine users' short-term interests.

\begin{figure*}[t!]
    \centering
    \includegraphics[width=0.9\linewidth]{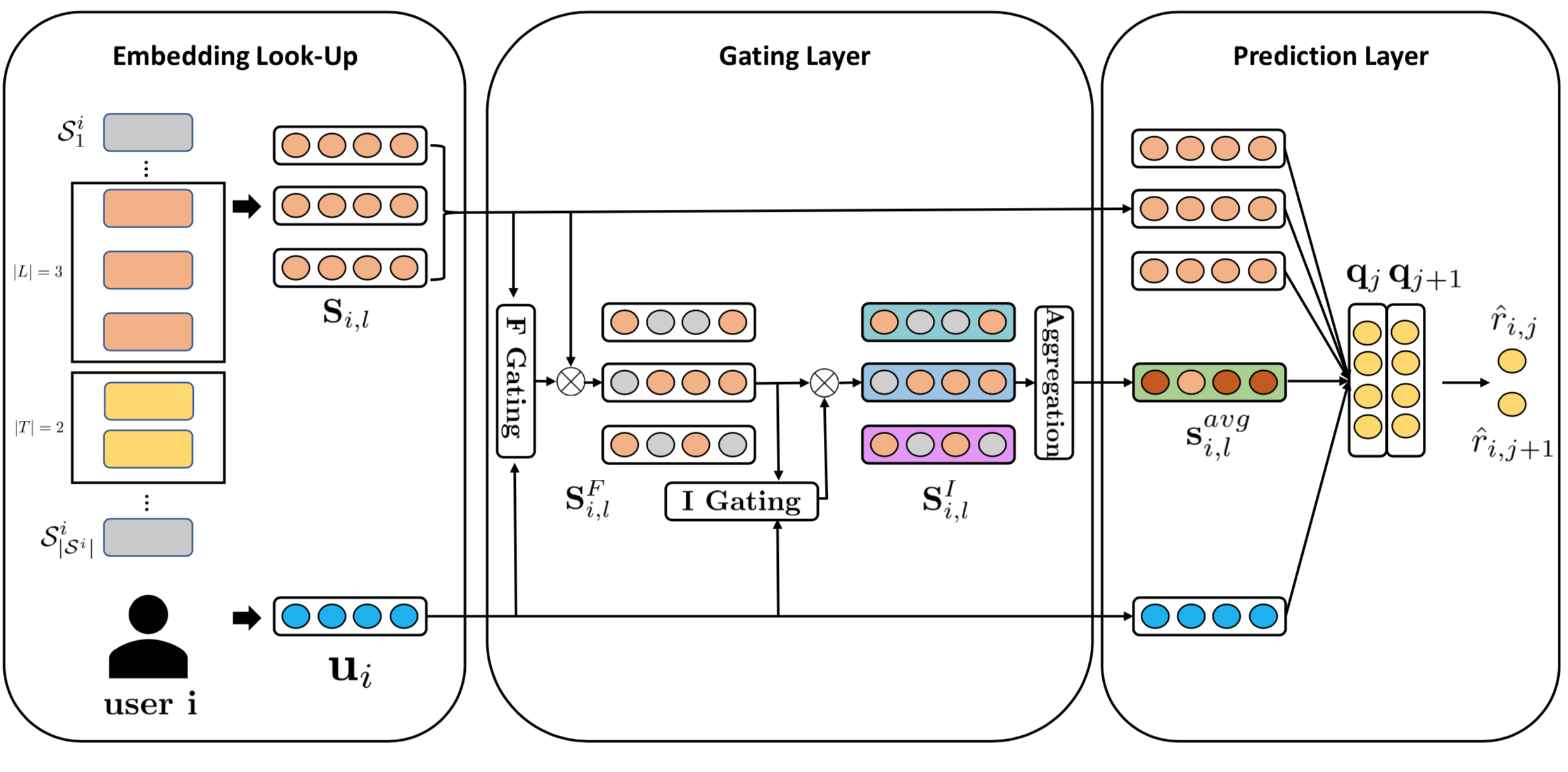}
    \caption{\label{fig:whole_model}The architecture of HGN. HGN consists of three major components: the embedding layer, the hierarchical gating layer, and the prediction layer. Specifically, \textit{F Gating} denotes the feature gating module, \textit{I Gating} denotes the instance gating module, \textit{Aggregation} denotes the aggregation layer, and $ \otimes $ denotes the element-wise multiplication.}
\end{figure*}

In this section, we introduce the proposed model to capture both the long-term interests and short-term interests of users for the sequential recommendation, which is shown in Figure \ref{fig:three_modules} and Figure \ref{fig:whole_model}. We first illustrate the hierarchical gating network for learning users' group-level preferences. Next, we present the inner product of item embeddings to model the item-item relations. Then we introduce the prediction layer for aggregating the long-term and short-term interests of users. Lastly, we go through the loss function and training process of the proposed model. 

\subsection{Hierarchical Gating for Group-level Influence}
In the sequential recommendation, taking advantage of the properties of sequential data to learn the (sub)sequence representation is a critical point, where an item may be closely related to its previous or subsequent items, or a group of previous items will have an impact on the items in the near future. In previous works, researchers have utilized various methods to model the group-level sequential interactions, e.g., convolutional neural networks \cite{DBLP:conf/wsdm/TangW18}, recurrent neural networks \cite{DBLP:journals/corr/HidasiKBT15,DBLP:conf/cikm/PeiYSZBT17,DBLP:conf/recsys/QuadranaKHC17,DBLP:conf/cikm/HidasiK18}, and the self-attention model \cite{DBLP:conf/icdm/KangM18}. Different from previous works, we propose a hierarchical gating network for modeling group-level user-item interactions, which consists of two components: a feature gating module and an instance gating module. These two modules allow the selection of effective latent features and relevant items, respectively, for predicting the subsequent items. Our proposed gating network is both effective and efficient (section \ref{sec:evaluation}).

\subsubsection{Feature Gating}
Unlike previous works \cite{DBLP:conf/wsdm/TangW18,DBLP:conf/cikm/HidasiK18,DBLP:conf/icdm/KangM18} that only operate on the item-level, we provide a learnable feature gating module to select salient latent features of items from the feature-level. For a certain item, some parts of the latent features are more relevant to predict the subsequent items. For example, for a big fan of Robert Downey Jr., after watching Iron Man I and Iron Man II, it is better to recommend Iron Man III rather than Aquaman, although Aquaman is also a superhero movie. Thus, to capture the representative item features based on users' long-term preferences is a necessary point to capture.

\textbf{Embedding Layer}. In the proposed module, the input is a sequence of $ |L| $ items, where each item is represented by a unique index. At the embedding layer, the item index is converted into a low-dimensional real-valued dense vector representation by an item embedding matrix $ \mathbf{E} \in \mathbb{R}^{d \times N} $, where $ d $ is the dimension of the item embedding and $ N $ is the number of items. After converted by the embedding layer, the item subsequence embeddings are represented as:
\[ 
\mathbf{S}_{i,l} = \begin{bmatrix}
 & | & | & | & \\ 
... & \mathbf{e}_{j-1} & \mathbf{e}_{j} & \mathbf{e}_{j+1} & ... \\ 
 & | & | & | & 
\end{bmatrix}
\]
where $ \mathbf{S}_{i,l} \in \mathbb{R}^{d \times |L|} $ indicates the embeddings of the $ l $-th subsequence of user $ i $, $ \mathbf{e}_{j} \in \mathbb{R}^{d} $ is the $ j $-th column of the embedding matrix $ \mathbf{E} $.

\textbf{Gated Linear Unit}. Inspired by the gated linear unit (GLU) proposed by Dauphin et al. in \cite{DBLP:conf/icml/DauphinFAG17}, which is utilized to control what information should be propagated for predicting the next word in the language modeling task, we also adopt a similar model to select what features are relevant to predict future items. The GLU in the original paper is shown:
\[ (\mathbf{X} \ast \mathbf{W} + \mathbf{b}) \otimes \sigma(\mathbf{X} \ast \mathbf{V} + \mathbf{c}), \] 
where $ \mathbf{X} $ is the input embeddings, $ \mathbf{W} $, $ \mathbf{V} $, $ \mathbf{b} $, $ \mathbf{c} $ are learnable parameters, $ \sigma $ is the sigmoid function, $ \ast $ is the convolution operation, and $ \otimes $ is the element-wise product between matrices.

\textbf{Personalized Feature Gating}. However, directly applying the GLU to select item features does not explicitly consider the user preference on items. For a certain item, a user may just focus a specific part of the item and neglect other unattractive parts. For example, a user may only care about whether the starring role is Tom Cruise rather than the movie content. 

Therefore, to capture the item features that tailored to users' preferences, we need to modify the GLU to be user-specific. To reduce the number of learnable parameters, we apply the inner product instead of the convolution operation in the original GLU (the superscript $ F $ indicates the item sequence embeddings are learned from the feature gating module):
\begin{equation}
\mathbf{S}_{i,l}^{F} = \mathbf{S}_{i,l} \otimes \sigma(\mathbf{W}_{g_1} \cdot \mathbf{S}_{i,l} + \mathbf{W}_{g_2} \cdot \mathbf{u}_{i} + \mathbf{b}_{g}),
\end{equation}
where $ \mathbf{u}_{i} \in \mathbb{R}^{d} $ is the embedding of user $ i $, $ \mathbf{W}_{g_1}, \mathbf{W}_{g_2} \in \mathbb{R}^{d \times d} $ and $ \mathbf{b}_g \in \mathbb{R}^{d} $ are learnable parameters, and $ \otimes $ is the element-wise product between matrices. By doing this, user-specific features of items can be passed to downstream layers.

\subsubsection{Instance Gating}
\textbf{Personalized Instance Gating}. Since our formulated problem is: given $ |L| $ successive items, how likely other items will appear after $ L $ in the near feature, we argue that there are some items are more relevant in $ L $ to predict the items users will interact. However, existing works either do not consider the representative items in $ L $ \cite{DBLP:conf/wsdm/TangW18,DBLP:conf/cikm/HidasiK18} or apply attention models to capture the representative items \cite{DBLP:conf/cikm/LiRCRLM17,DBLP:conf/icdm/KangM18}. Unlike previous works benefiting from attention models, we adopt an instance-level gating module to select the informative items that are helpful to predict items in the near future according to users' preferences:
\begin{equation}
\mathbf{S}^{I}_{i,l} = \mathbf{S}^{F}_{i,l} \otimes \sigma(\mathbf{w}_{g_3}^{\top} \cdot \mathbf{S}^{F}_{i,l} + \mathbf{u}_{i}^{\top} \cdot \mathbf{W}_{g_4}),
\end{equation}
where $ \mathbf{S}^{I}_{i,l} \in \mathbb{R}^{d \times |L|} $ is the sequence embedding after the instance gating, $ \mathbf{w}_{g_3} \in \mathbb{R}^{d} $, $ \mathbf{W}_{g_4} \in \mathbb{R}^{d \times |L|} $ are learnable parameters. By applying the instance gating, the representative items will contribute more to make predictions about the future items and irrelevant items will be largely neglected.

\textbf{Aggregation Layer}. To make the item embeddings $ \mathbf{S}^{I}_{i,l} $ into one group-level latent representation, we can either apply average pooling or max pooling on $ \mathbf{S}^{I}_{i,l} $:
\begin{equation}
\mathbf{s}_{i,l}^{avg} = avg-pooling(\mathbf{S}^{I}_{i,l}),
\end{equation}
\begin{equation}
\mathbf{s}_{i,l}^{max} = max-pooling(\mathbf{S}^{I}_{i,l}),
\end{equation}
where $ \mathbf{s}_{i,l}^{avg}, \mathbf{s}_{i,l}^{max} \in \mathbb{R}^{d} $. Since the item embeddings have manipulated by the feature-level and instance-level gating modules, the informative features and items have been selected and irrelevant ones have been eliminated. Thus, the average pooling will accumulate the informative parts in these embeddings. On the other hand, max-pooling directly selects the most representative features from each embedding to form the group-level representation.

\subsection{Item-item Product}
The relation between two single items is an important factor to model in the recommendation task and has been widely studied in many years \cite{DBLP:reference/sp/NingDK15,DBLP:conf/kdd/KabburNK13}, e.g., item-based collaborative filtering methods utilizing the rating vectors of two items to calculate the similarity. However, most of the recent works \cite{DBLP:conf/wsdm/TangW18,DBLP:conf/cikm/HidasiK18,DBLP:conf/icdm/KangM18} only consider the sequential recommendation from the group-level, but do not explicitly capture the item-item relations between the items in $ L $ and the items user will interact in the future. Since strongly related item pairs will appear in $ L $ and $ T $ simultaneously. Unlike previous works, we apply the inner product between the input item embeddings and the output item embeddings to capture the item relations between $ L $ and $ T $:
\[
\sum_{\mathbf{e}_j \in \mathbf{S}_{i,l}} \mathbf{e}^{\top}_j \cdot \mathbf{Q},
\]
where $ \mathbf{Q} \in \mathbb{R}^{d \times N} $ is the output item embeddings, the sum of multiplication results captures the accumulated item-item relation scores from each item in $ L $ to all other items.

\subsection{Prediction Layer}
After applying the hierarchical gating network to capture the short-term interests of users and item-item product to capture the relevant item pairs, we adopt the classical matrix factorization term to capture the global and long-time interests of users. Given the $ l $-th subsequence to predict, the prediction score of user $ i $ on item $ j $ is:
\begin{equation}
\hat{r}_{i,j} = \mathbf{u}_{i}^{\top} \cdot \mathbf{q}_{j} + \mathbf{s}^{avg \top}_{i,l} \cdot \mathbf{q}_{j} + \sum_{\mathbf{e}_{k} \in \mathbf{S}_{i,l}} \mathbf{e}^{\top}_{k} \cdot \mathbf{q}_{j},
\end{equation}
where $ \mathbf{q}_{j} \in \mathbb{R}^{d} $ is the $ j $-th column of the output item embedding $ \mathbf{Q} $. In the prediction layer, the first term captures the user long-term interests, the second term models the user short-term interests, and the third term reflects the relations between item pairs.

\subsection{Network Training} \label{sec:network training}
As the training data is from the user implicit feedback, we optimize the proposed model by the Bayesian Personalized Ranking objective \cite{DBLP:conf/uai/RendleFGS09}: optimizing the pairwise ranking between the positive and non-observing items:
\begin{equation}
\operatorname*{arg\,min}_{\mathbf{U}, \mathbf{Q}, \mathbf{E}, \mathbf{\Theta}} \sum_{(i, L_i, j, k) \in \mathcal{D}} -log\sigma(\hat{r}_{i,j} - \hat{r}_{i,k}) + \lambda(||\mathbf{U}||^{2} + ||\mathbf{Q}||^{2} + ||\mathbf{E}||^{2} + ||\mathbf{\Theta}||^{2}),
\end{equation}
where $ L_i $ denotes one of the $ |L| $ successive items of user $ i $, $ j $ denotes the item that in $ T_i $, and $ k $ denotes the randomly sampled negative item, $ \mathbf{\Theta} $ is the parameters in the gating network, $ \lambda $ is the regularization parameter. By minimizing the objective function, the partial derivatives with respect to all the parameters can be computed by gradient descent with back-propagation. We apply Adam \cite{DBLP:journals/corr/KingmaB14} to automatically adapt the learning rate during the learning procedure. 

\textbf{Time complexity}. The computational complexity of our model for each $ L $ is mainly due to the feature gating layer and item-item product module, which is $ O(|L|d^2 + |L|Nd) $ ($ |L| $ is the length of $ L $, $ d $ is the dimension of embeddings, and $ N $ is the number of items). This computational complexity makes our model scalable on large datasets. We empirically test the training speed with other state-of-the-art methods and find that our model is faster than other methods (section \ref{sec:training_efficiency}).

\section{Experiments} \label{sec:evaluation}
In this section, we evaluate the proposed model with the state-of-the-art methods on five real-world datasets\footnote{The code is available on Github: https://github.com/allenjack/HGN}.

\subsection{Datasets}
The proposed model is evaluated on five real-world datasets from various domains with different sparsities: \textit{MovieLens-20M} \cite{DBLP:journals/tiis/HarperK16}, \textit{Amazon-Books} and \textit{Amazon-CDs} \cite{DBLP:conf/www/HeM16}, \textit{Goodreads-Children} and \textit{Goodreads-Comics} \cite{DBLP:conf/recsys/WanM18}. \textit{MovieLens-20M} is a user-movie dataset collected from the \textit{MovieLens} website, where this dataset has 20 million user-movie interactions. The \textit{Amazon-Books} and \textit{Amazon-CDs} datasets are adopted from the Amazon review dataset\footnote{http://jmcauley.ucsd.edu/data/amazon/} with different categories, i.e., CDs and Books, which cover a large amount of user-item interaction data, e.g., user ratings and reviews. \textit{Goodreads-Children} and \textit{Goodreads-Comics} datasets\footnote{https://sites.google.com/eng.ucsd.edu/ucsdbookgraph/home} are collected in late 2017 from \textit{goodreads} website with different genres, and we use the genres of Children and Comics. In order to be consistent with the implicit feedback setting, we keep those with ratings no less than four (out of five) as positive feedback and treat all other ratings as missing entries on all datasets. To filter noisy data, we only keep the users with at least ten ratings and the items at least with five ratings. The data statistics after preprocessing are shown in Table \ref{tab:data_statistics}. 

For each user, we hold the 70\% of interactions in the user sequence as the training set and use the next 10\% of interactions as the validation set for hyper-parameter tuning. The remaining 20\% constitutes the test set for reporting model performance. Note that during the testing procedure, the input sequences include the interactions in both the training set and validation set. The execution of all the models is carried out five times independently, and we report the average results.

\begin{table}[ht]
\centering
\caption{\label{tab:data_statistics}The statistics of datasets.}
\begin{tabular}{ |c|c|c|c|c|c| }
 \hline
 Dataset & \#Users & \#Items & \#Interactions & Density \\
 \hline
 \textit{ML20M} & 129,797 & 13,649 & 9,921,393 & 0.560\% \\ 
 \hline
 \textit{Books} & 52,406 & 41,264 & 1,856,747 & 0.086\% \\ 
 \hline
 \textit{CDs} & 17,052 & 35,118 & 472,265 & 0.079\% \\ 
 \hline
 \textit{Children} & 48,296 & 32,871 & 2,784,423 & 0.175\% \\ 
 \hline
 \textit{Comics} & 34,445 & 33,121 & 2,411,314 & 0.211\% \\ 
 \hline
\end{tabular}
\vspace{-0.3cm}
\end{table}

\subsection{Evaluation Metrics}
We evaluate our model versus other methods in terms of \textit{Recall@k} and \textit{NDCG@k}. For each user, Recall@k (R@k) indicates what percentage of her rated items can emerge in the top $ k $ recommended items. NDCG@k (N@k) is the normalized discounted cumulative gain at $ k $, which takes the position of correctly recommended items into account. 

\subsection{Methods Studied}
To demonstrate the effectiveness of our model, we compare to the following recommendation methods.

\textit{Classical methods for implicit feedback}:
\begin{itemize}
\item \textbf{BPRMF}, the Bayesian Personalized Ranking based matrix factorization \cite{DBLP:conf/uai/RendleFGS09}, which is a classic method for learning pairwise personalized rankings from user implicit feedback. Specifically, we use BPR-MF for model learning.
\end{itemize}

\textit{State-of-the-art session-based recommendation methods}:
\begin{itemize}
\item \textbf{GRU4Rec}, gated recurrent unit for recommendation \cite{DBLP:journals/corr/HidasiKBT15}, which uses recurrent neural networks to model user-item interaction sequences for session-based recommendation. Each user sequence is treated as a session.
\item \textbf{GRU4Rec+}, an improved version of GRU4Rec \cite{DBLP:conf/cikm/HidasiK18}, which adopts a different loss function and sampling strategy, and shows significant performance gains on Top-N recommendation.
\item \textbf{NextItNet}, the next item recommendation net \cite{DBLP:conf/wsdm/YuanKAJ019}, applies dilated convolutional neural networks to increase the receptive fields without relying on the pooling operation.
\end{itemize}

\textit{State-of-the-art sequential recommendation methods}:
\begin{itemize}
\item \textbf{Caser}, convolutional sequence embedding model \cite{DBLP:conf/wsdm/TangW18}, which captures high-order Markov chains by applying convolution operations on the embeddings of the $ |L| $ recent items.
\item \textbf{SASRec}, self-attention based sequential model \cite{DBLP:conf/icdm/KangM18}, which uses an attention mechanism to identify relevant items for predicting the next item.
\end{itemize}

\textit{The proposed method}:
\begin{itemize}
\item \textbf{HGN}, the proposed model, applies a hierarchical gating network to learn the group-level representations of a sequence of items and adopts the item-item product to explicitly capture the item-item relations.
\end{itemize}

Given our extensive comparisons against the state-of-the-art methods, we omit comparisons with methods such as FMC and FPMC \cite{DBLP:conf/www/RendleFS10}, Fossil \cite{DBLP:conf/icdm/HeM16}, since they have been outperformed by the recently proposed Caser and SASRec.

\subsection{Experiment Settings}
In the experiments, the latent dimension of all the models is set to 50. For those session-based methods, we treat each user sequence as one session. For GRU4Rec and GRU4Rec+, we find that when the learning rate is $ 0.001 $, and batch size is $ 50 $ can achieve good performance. These two methods adopt Top1 loss and BPR-max loss, respectively. For NextItNet, we following the original settings in the paper to set the learning rate to $ 0.001 $, the kernel size to $ 3 $, the dilated levels to $ 1 $ and $ 2 $, the batch size to $ 32 $. For Caser, we follow the settings in the author-provided code to set $ |L|=5 $, $ |T|=3 $, the number of horizontal filters to $ 16 $, the number of vertical filters to $ 4 $, where Caser can achieve good results. For SASRec, we set the number of self-attention blocks to $ 2 $, the batch size to $ 128 $, and the maximum sequence length to $ 50 $. The network architectures of above methods are also set the same with the original papers. The hyper-parameters are tuned using the validation set.

For HGN, we follow the same setting in Caser to set $ |L|=5 $ and $ |T|=3 $, where the length effects are shown in the section \ref{subsec:parameter_sensitivity}. Hyper-parameters are tuned by grid search on the validation set. The network embedding size $ d $ is also set to $ 50 $. The learning learning rate and $ \lambda $ are set to $ 0.001 $ and $ 0.001 $, respectively. The batch size is set to $ 4096 $. Our experiments are conducted with PyTorch\footnote{https://pytorch.org/} running on GPU machines (Nvidia GeForce GTX 1080 Ti).

\subsection{Performance Comparison}
\begin{figure}[t!]
    \centering
    \begin{subfigure}[t]{0.25\textwidth}
        \centering
        \includegraphics[width=\linewidth]{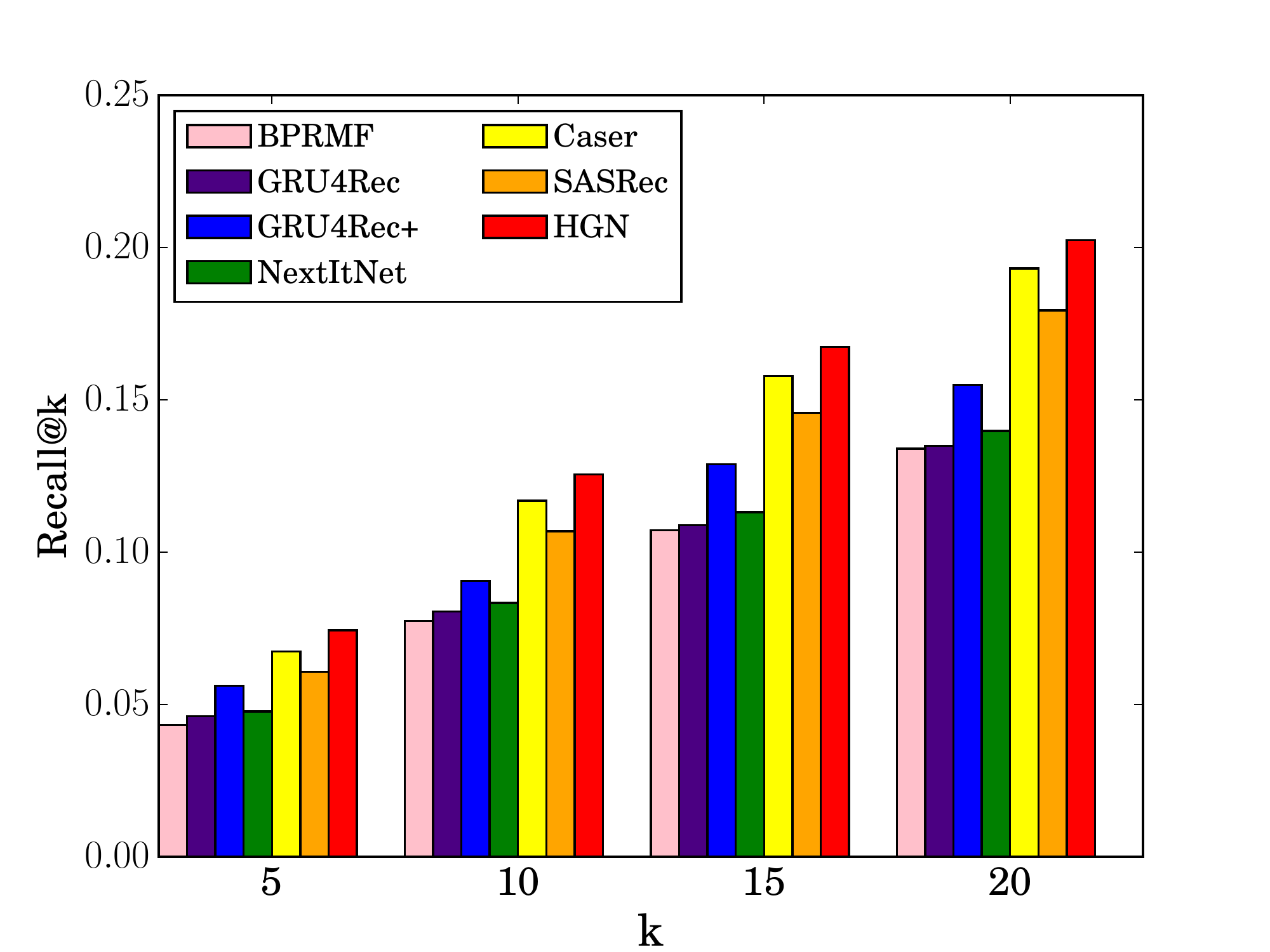}
        \caption{\label{fig:ml20m_recall} Recall@k on MovieLens-20M}
    \end{subfigure}%
    \begin{subfigure}[t]{0.25\textwidth}
        \centering
        \includegraphics[width=\linewidth]{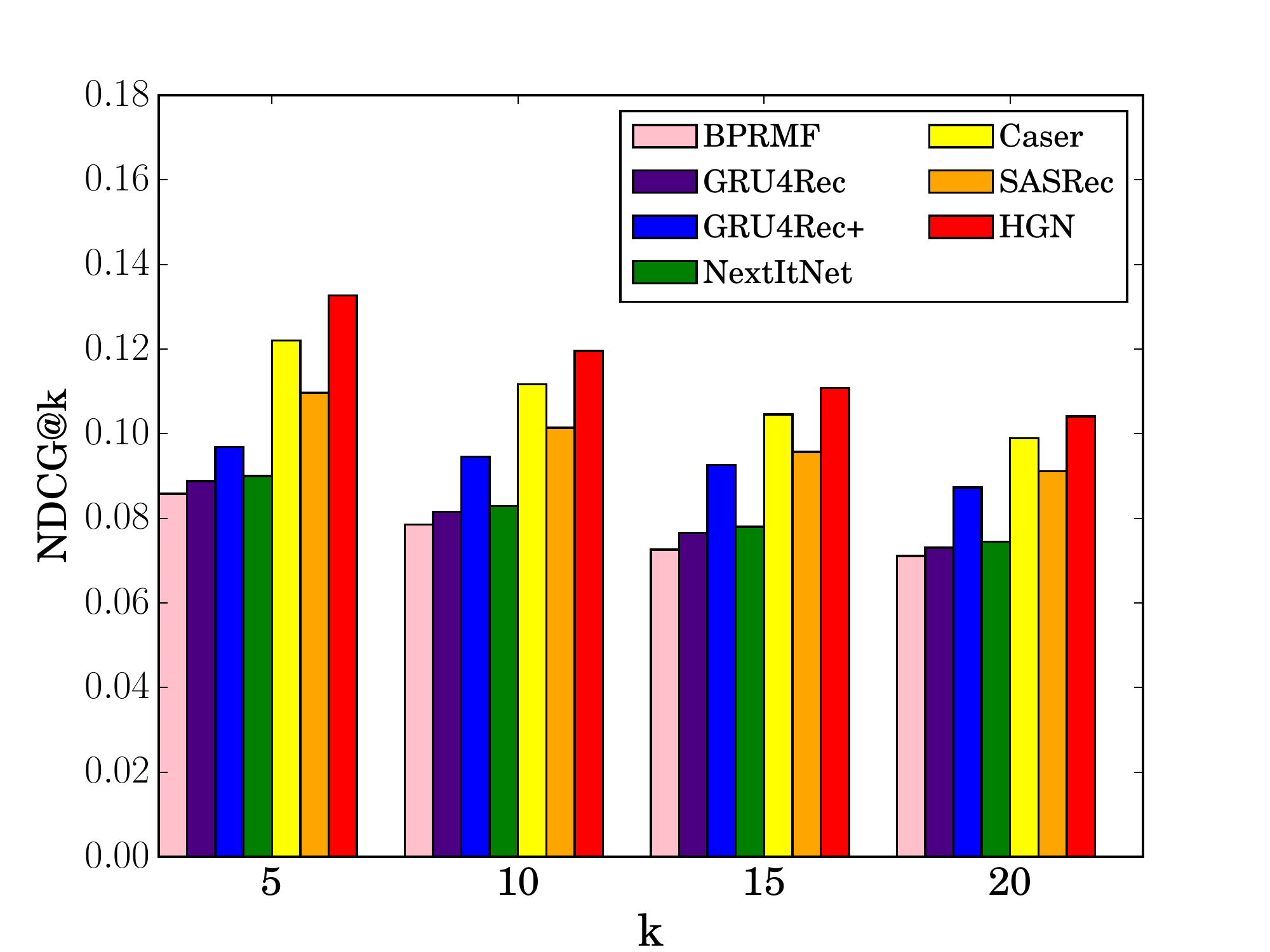}
        \caption{\label{fig:ml20m_ndcg} NDCG@k on MovieLens-20M}
    \end{subfigure}
    \caption{\label{fig:ml20m}The performance comparison on MovieLens-20M.}
\vspace{-0.3cm}
\end{figure}

\begin{figure}[t!]
    \centering
    \begin{subfigure}[t]{0.25\textwidth}
        \centering
        \includegraphics[width=\linewidth]{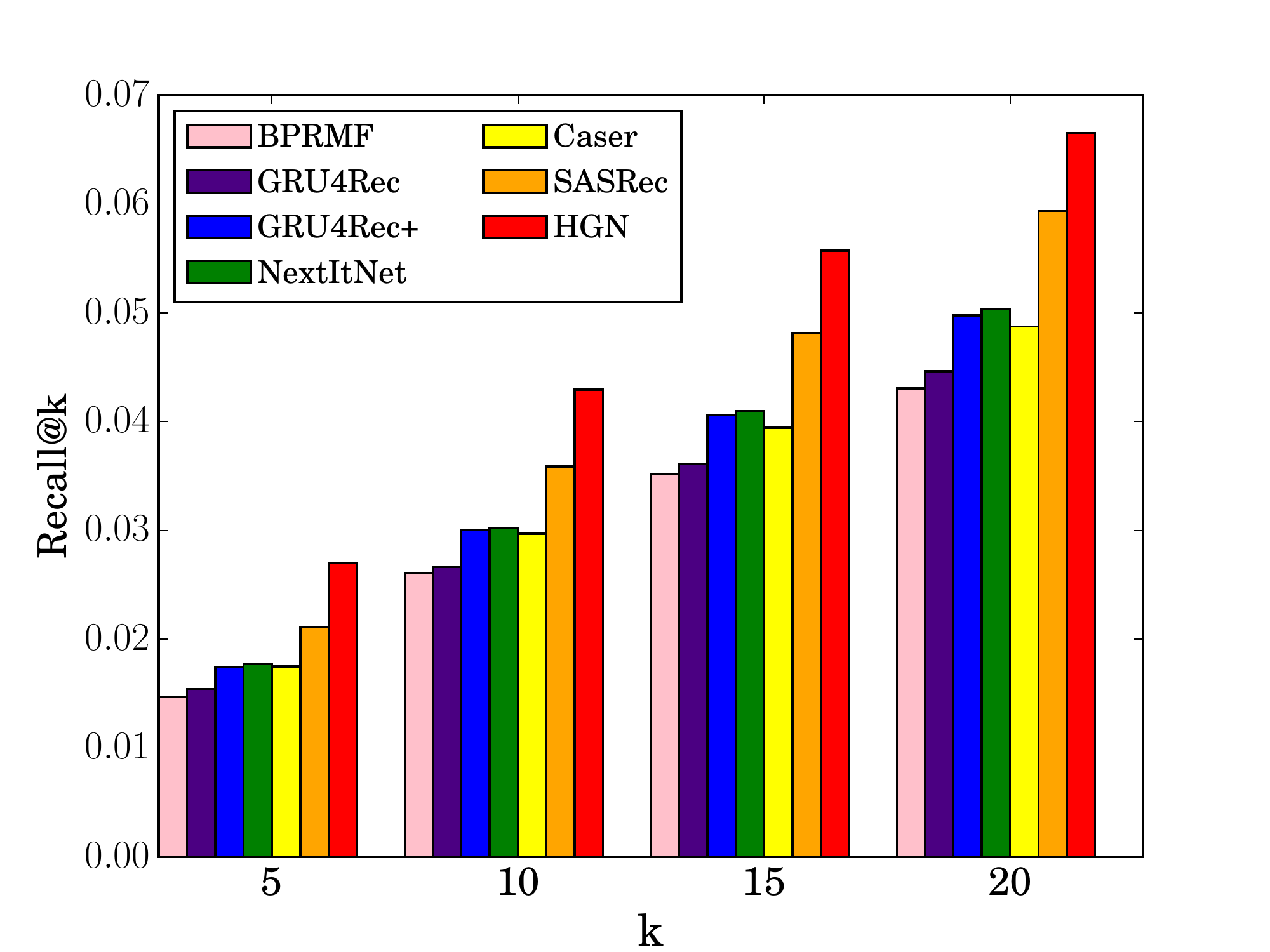}
        \caption{\label{fig:Books_recall} Recall@k on Amazon-Books}
    \end{subfigure}%
    \begin{subfigure}[t]{0.25\textwidth}
        \centering
        \includegraphics[width=\linewidth]{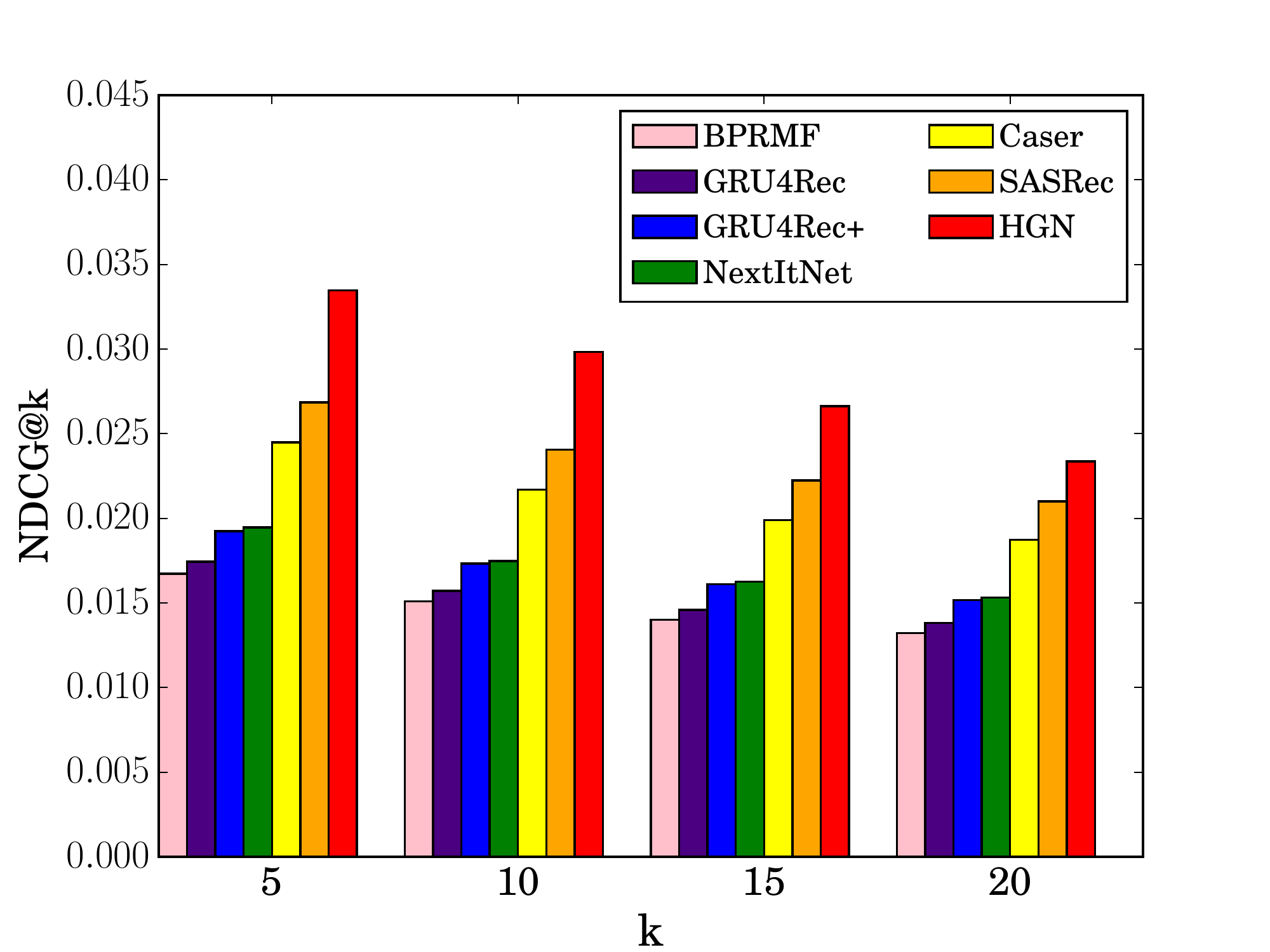}
        \caption{\label{fig:Books_ndcg} NDCG@k on Amazon-Books}
    \end{subfigure}
    \caption{\label{fig:books}The performance comparison on Amazon-Books.}
\vspace{-0.3cm}
\end{figure}

\begin{figure}[t!]
    \centering
    \begin{subfigure}[t]{0.25\textwidth}
        \centering
        \includegraphics[width=\linewidth]{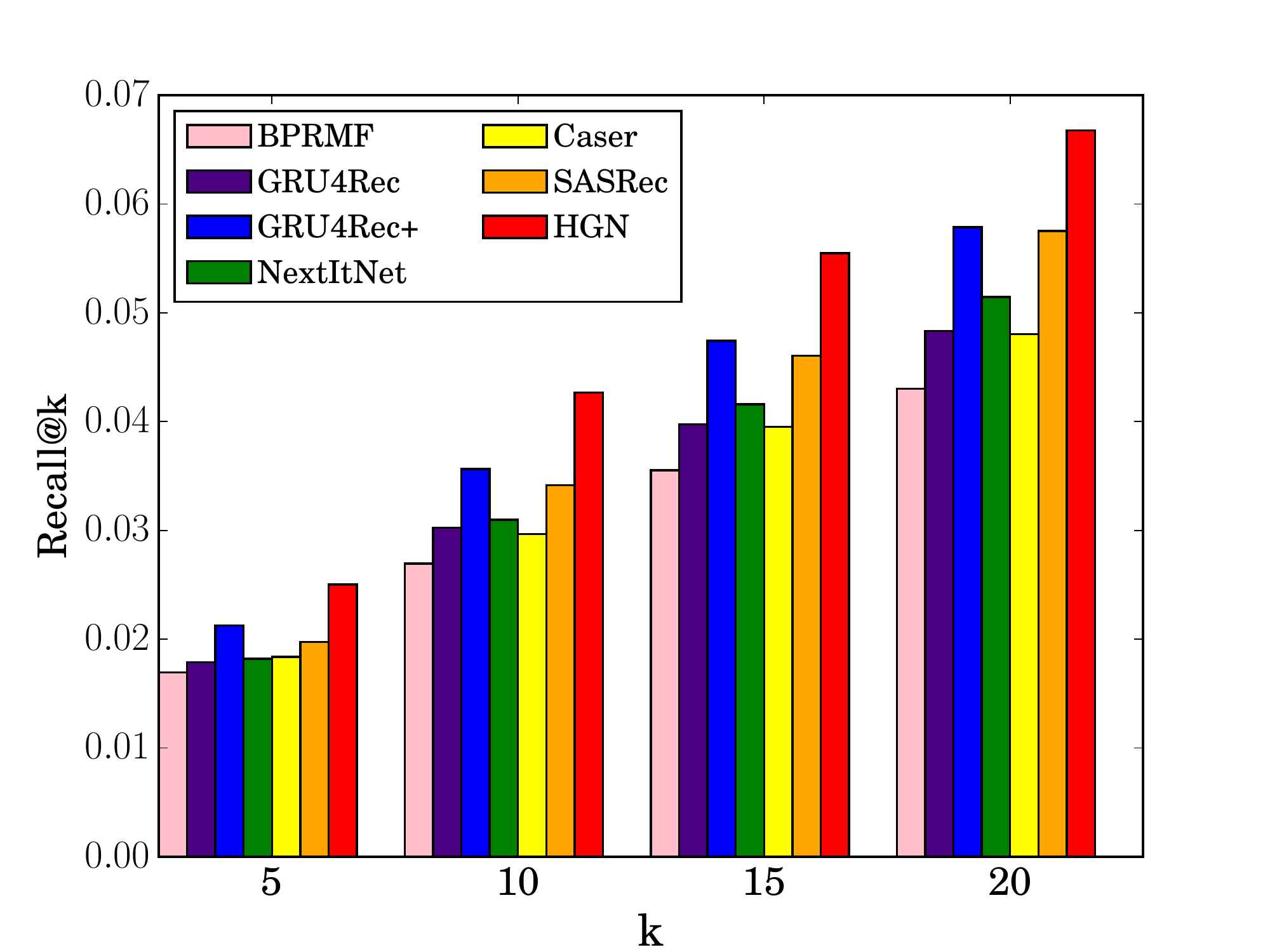}
        \caption{\label{fig:CDs_recall} Recall@k on Amazon-CDs}
    \end{subfigure}%
    \begin{subfigure}[t]{0.25\textwidth}
        \centering
        \includegraphics[width=\linewidth]{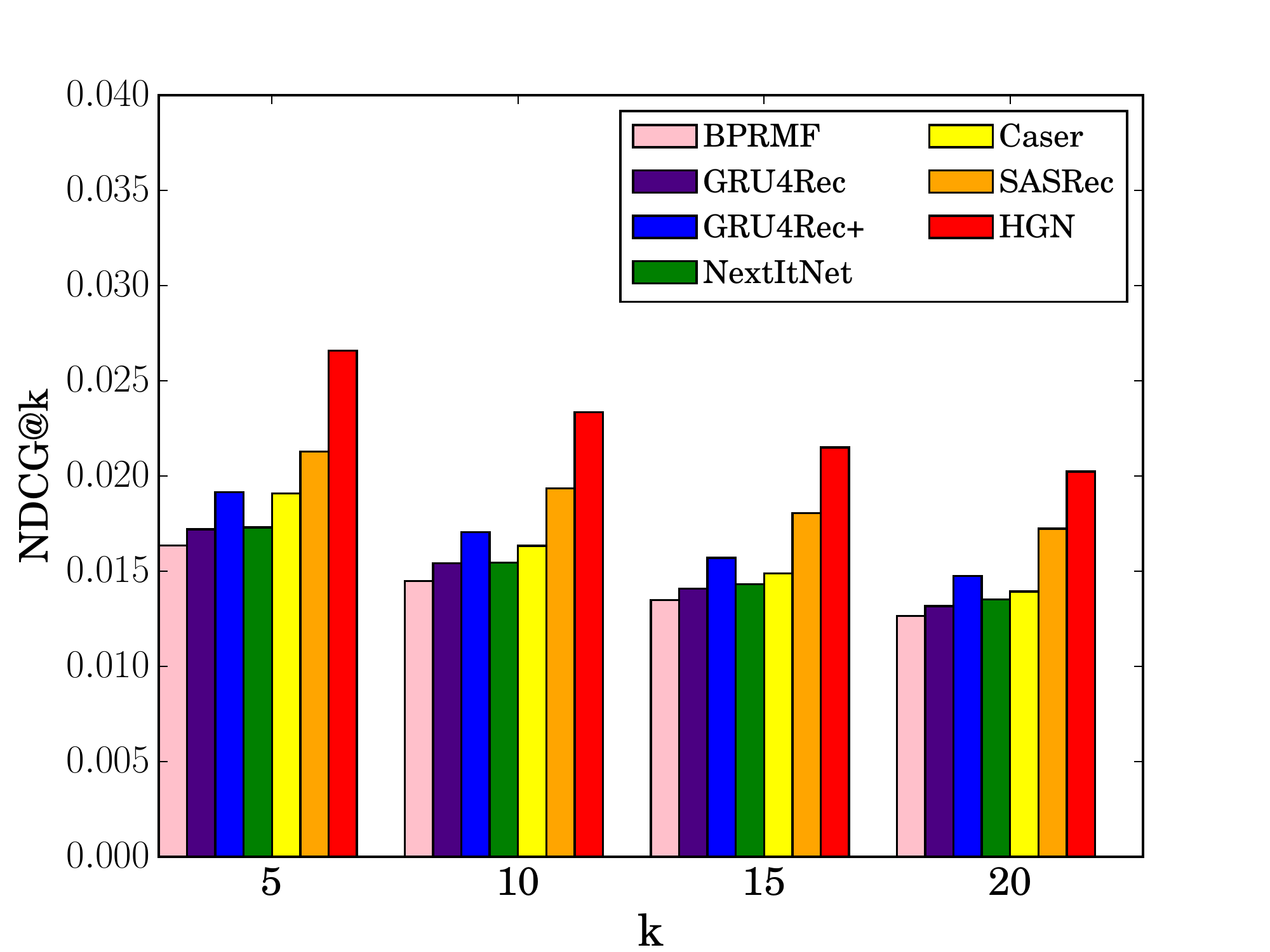}
        \caption{\label{fig:CDs_ndcg} NDCG@k on Amazon-CDs}
    \end{subfigure}
    \caption{\label{fig:cds}The performance comparison on Amazon-CDs.}
\vspace{-0.3cm}
\end{figure}

\begin{figure}[t!]
    \centering
    \begin{subfigure}[t]{0.25\textwidth}
        \centering
        \includegraphics[width=\linewidth]{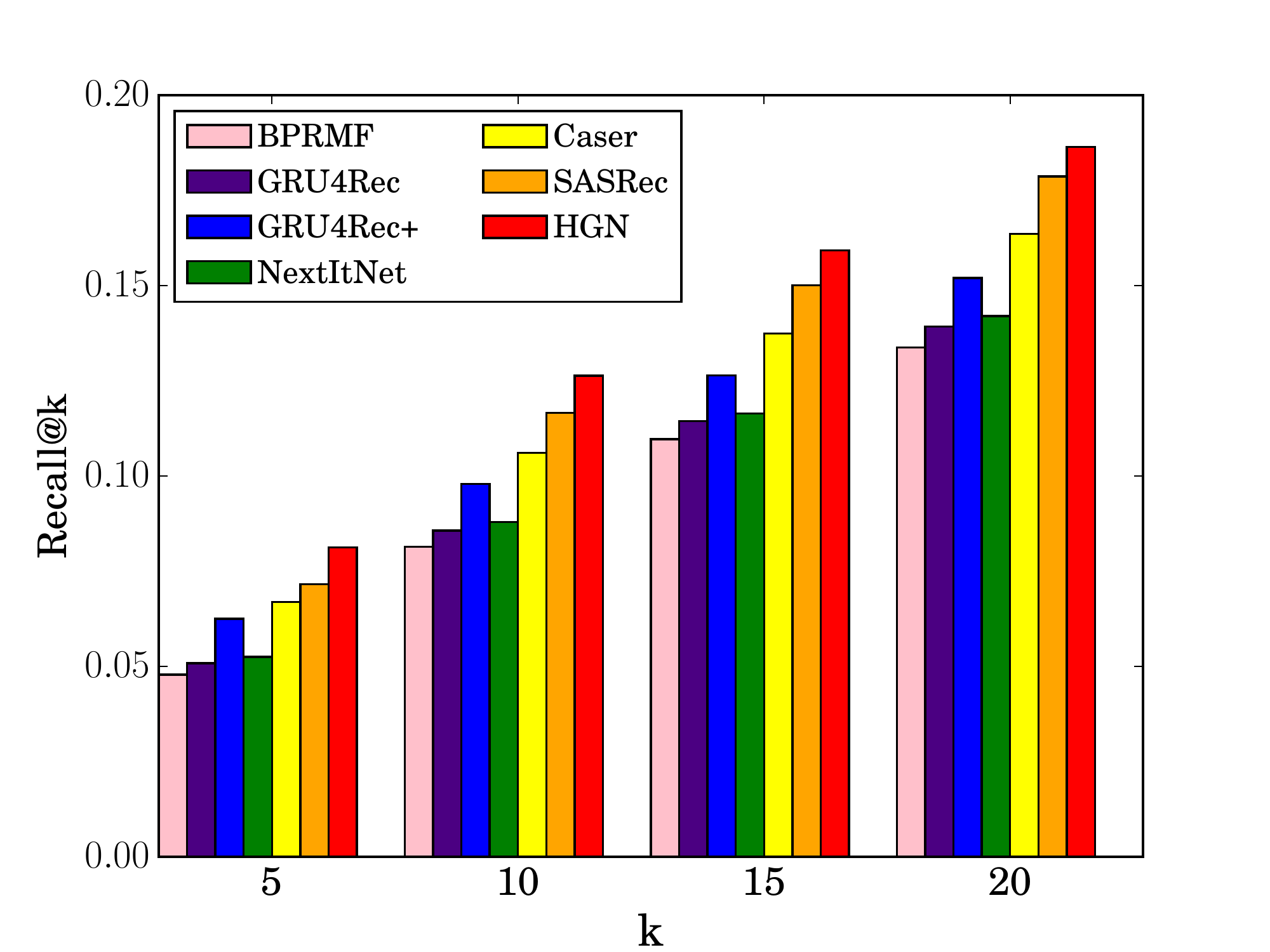}
        \caption{\label{fig:Children_recall} Recall@k on Children}
    \end{subfigure}%
    \begin{subfigure}[t]{0.25\textwidth}
        \centering
        \includegraphics[width=\linewidth]{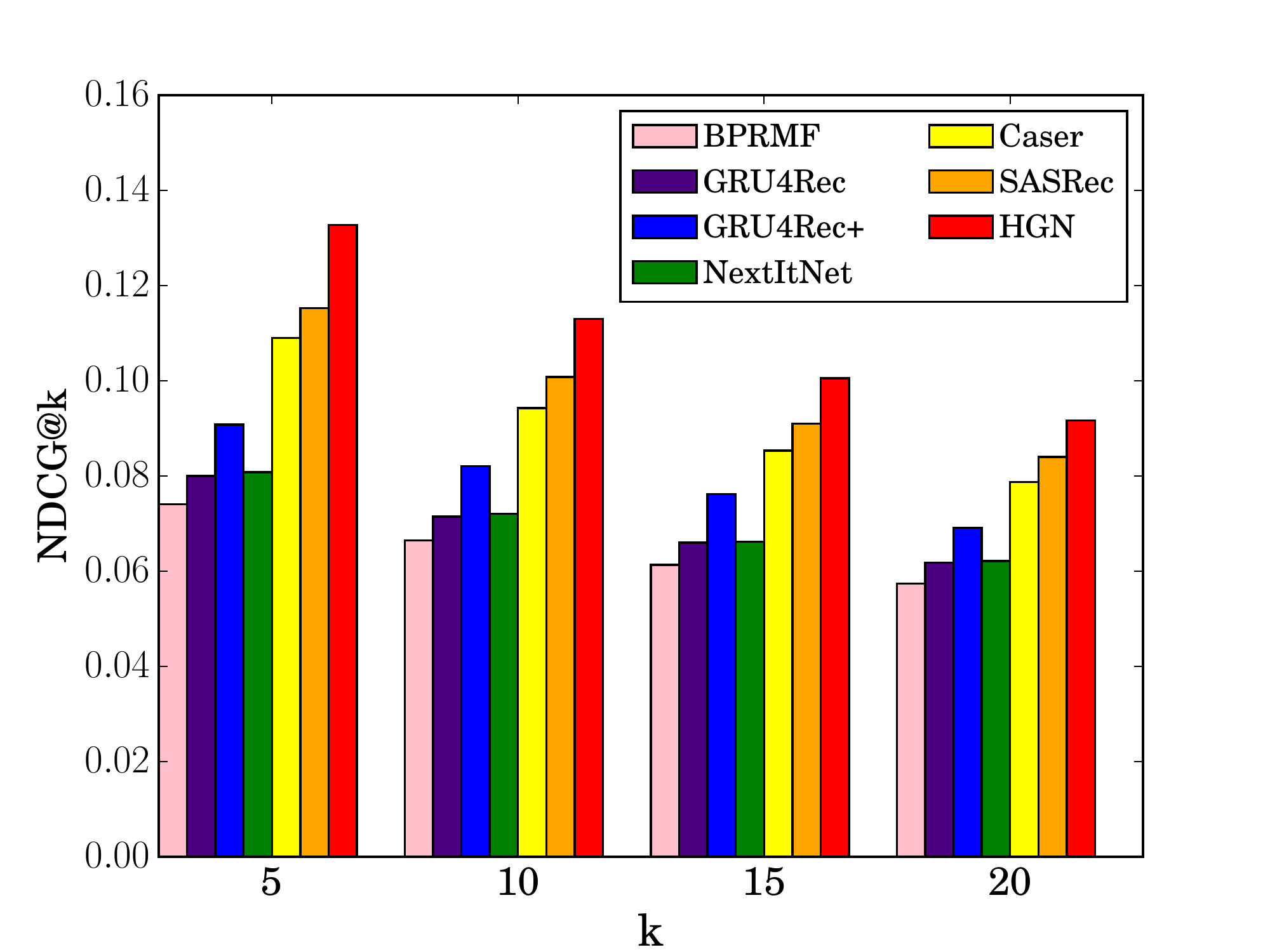}
        \caption{\label{fig:Children_ndcg} NDCG@k on Children}
    \end{subfigure}
    \caption{\label{fig:children}The performance comparison on Children.}
\vspace{-0.3cm}
\end{figure}

\begin{figure}[t!]
    \centering
    \begin{subfigure}[t]{0.25\textwidth}
        \centering
        \includegraphics[width=\linewidth]{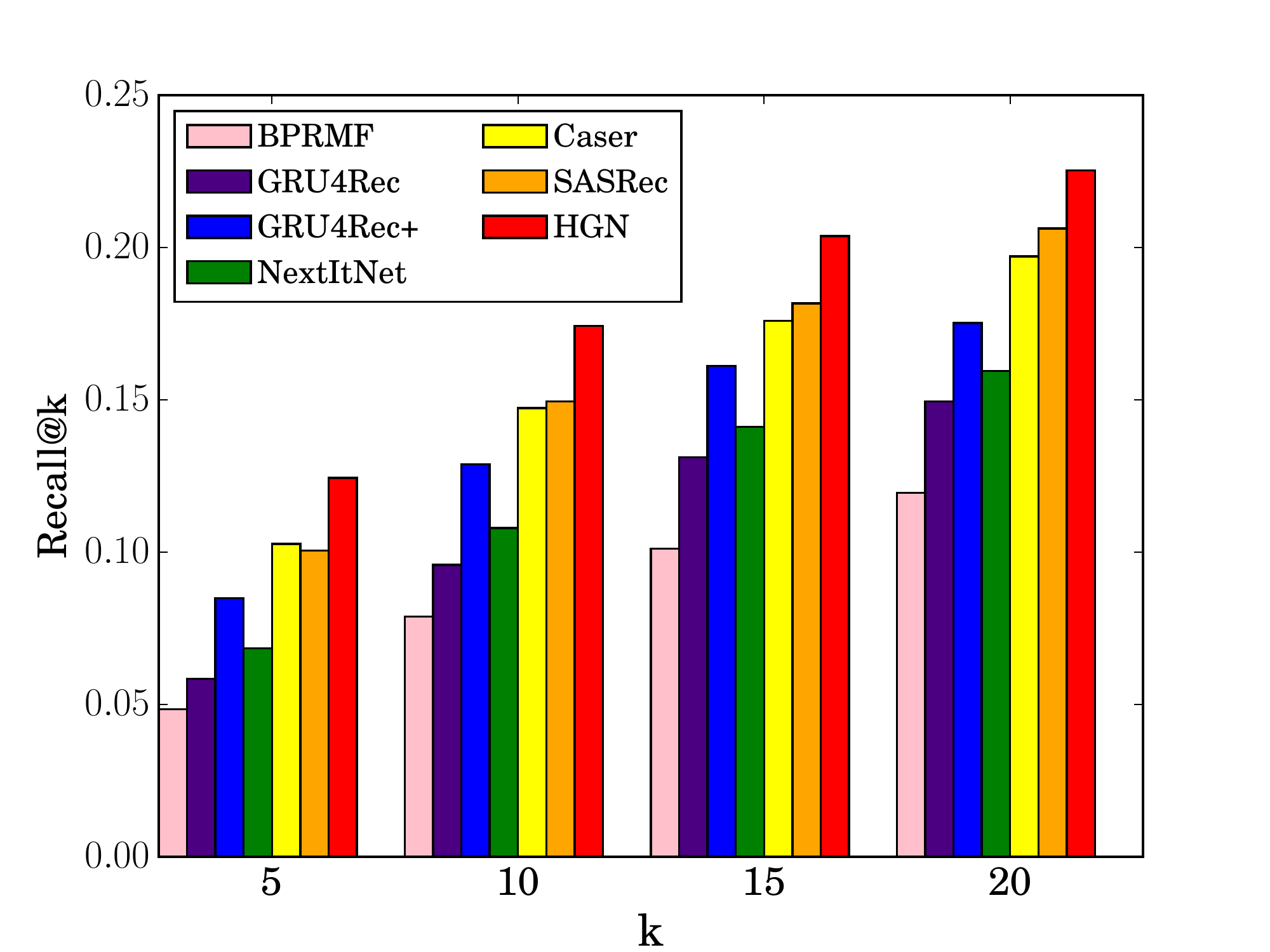}
        \caption{\label{fig:Comics_recall} Recall@k on Comics}
    \end{subfigure}%
    \begin{subfigure}[t]{0.25\textwidth}
        \centering
        \includegraphics[width=\linewidth]{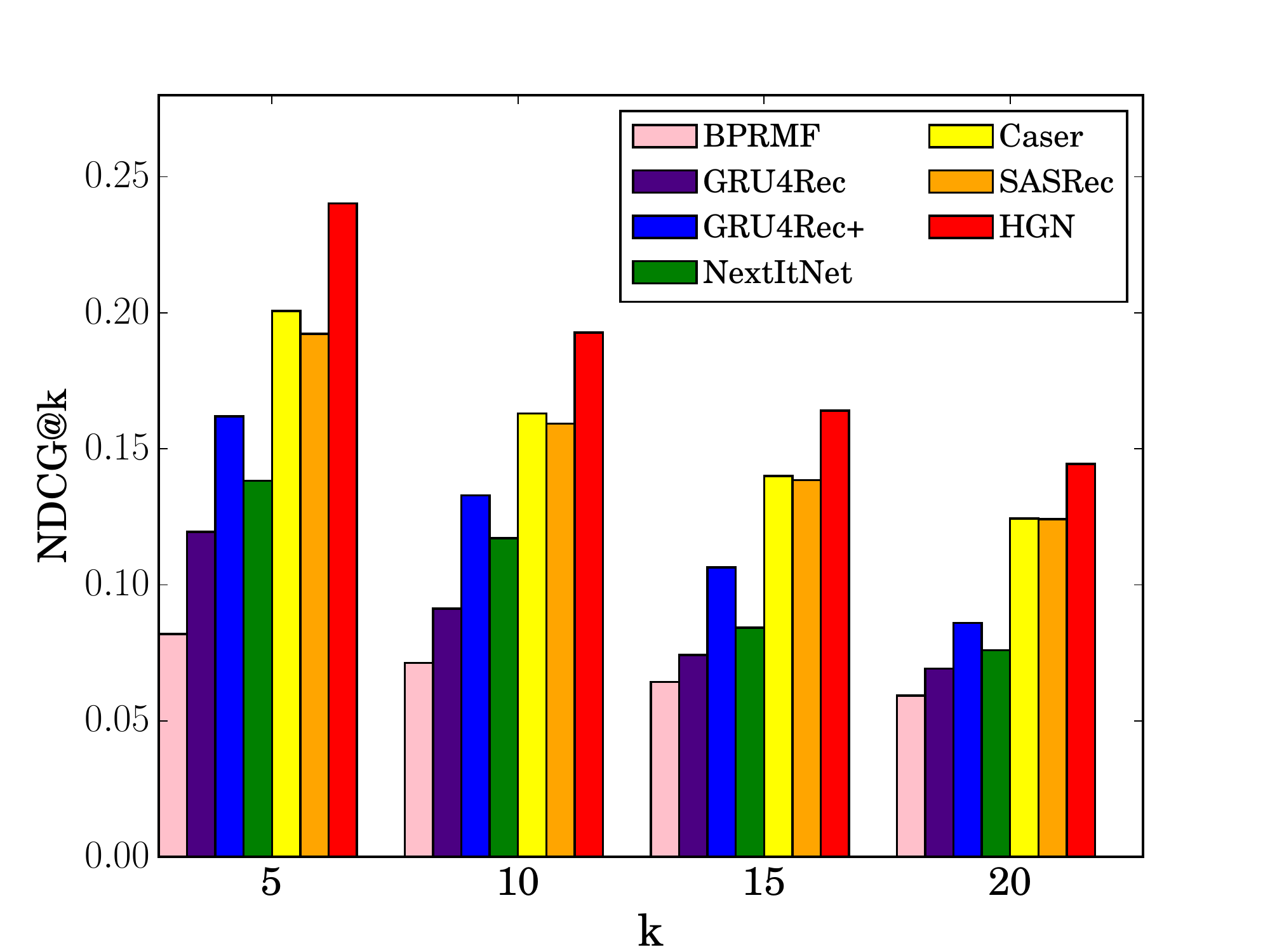}
        \caption{\label{fig:Comics_ndcg} NDCG@k on Comics}
    \end{subfigure}
    \caption{\label{fig:comics}The performance comparison on Comics.}
\vspace{-0.3cm}
\end{figure}

The performance comparison results are shown in Figure \ref{fig:ml20m}, \ref{fig:books}, \ref{fig:cds}, \ref{fig:children}, \ref{fig:comics}, and Table \ref{tab:performance_comparison}. 

\begin{table*}[ht]
\caption{\label{tab:performance_comparison}The performance comparison of all methods in terms of \textit{Recall@10} and \textit{NDCG@10}. The best performing method is boldfaced. The underlined number is the second best performing method. $ * $, $ ** $, $ *** $ indicate the statistical significance for $ p <= 0.05 $, $ p <= 0.01 $, and $ p <= 0.001 $, respectively, compared to the best baseline method based on the paired t-test. \textit{Improv.} denotes the improvement of our model over the best baseline method.}
\begin{tabular}{|c|c| c c c| c c| l| c|}
\hline
& \textbf{BPRMF} & \textbf{GRU4Rec} & \textbf{GRU4Rec+} & \textbf{NextItRec} & \textbf{Caser} & \textbf{SASRec} & \textbf{HGN} & \multicolumn{1}{l|}{\textbf{Improv.}} \\\hline
\multicolumn{9}{|c|}{Recall@10} \\
\hline
\textit{MovieLens-20M} & 0.0774 & 0.0804 & 0.0904 & 0.0833 & \underline{0.1169} & 0.1069 & \textbf{0.1255*} & 7.36\% \\
\textit{Amazon-Books} & 0.0260 & 0.0266 & 0.0301 & 0.0303 & 0.0297 & \underline{0.0358} & \textbf{0.0429***} & 19.83\% \\
\textit{Amazon-CDs} & 0.0269 & 0.0302 & \underline{0.0356} & 0.0310 & 0.0297 & 0.0341 & \textbf{0.0426**} & 19.66\% \\
\textit{GoodReads-Children} & 0.0814 & 0.0857 & 0.0978 & 0.0879 & 0.1060 & \underline{0.1165} & \textbf{0.1263*} & 8.41\% \\
\textit{GooReads-Comics} & 0.0788  & 0.0958 & 0.1288 & 0.1078 & 0.1473 & \underline{0.1494} & \textbf{0.1743***} & 16.67\% \\
\hline
\multicolumn{9}{|c|}{NDCG@10} \\ 
\hline
\textit{MovieLens-20M} & 0.0785 & 0.0815 & 0.0946 & 0.0828 & \underline{0.1116} & 0.1014 & \textbf{0.1195*} & 7.07\% \\
\textit{Amazon-Books} & 0.0151  & 0.0157 & 0.0173 & 0.0174 & 0.0216 & \underline{0.0240} & \textbf{0.0298***} & 24.17\% \\ 
\textit{Amazon-CDs} & 0.0145 & 0.0154 & 0.0171 & 0.0155 & 0.0163 & \underline{0.0193} & \textbf{0.0233**} & 20.73\% \\
\textit{GoodReads-Children} & 0.0664 & 0.0715 & 0.0821 & 0.0720 & 0.0943 & \underline{0.1007} & \textbf{0.1130*} & 12.21\% \\
\textit{GoodReads-Comics} & 0.0713 & 0.0912 & 0.1328 & 0.1171 & \underline{0.1629} & 0.1592 & \textbf{0.1927***} & 18.29\% \\
\hline
\end{tabular}
\vspace{-0.3cm}
\end{table*}

\textbf{Observations about our model}. First, the proposed model---HGN, achieves the best performance on five datasets with all evaluation metrics, which illustrates the superiority of our model. Second, HGN achieves better performance than SASRec. The reasons are three-fold: (1) SASRec only applies the instance-level selection but neglecting the feature-level one, which plays an important role in learning short-term user interests (section \ref{sec:ablation}); (2) SASRec adopts a hyper-parameter---the maximum sequence length to reduce the computation burden, where only using part of the user data may lead to the insufficient understanding of long-term user interests; (3) SASRec does not explicitly model the item-item relations between two closely relevant items, which is captured by our item-item product module. Third, HGN outperforms Caser, one major reason is that Caser only applies CNNs to learn the group-level representation of several successive items without considering the item importance for different users. Fourth, HGN obtains better results than GRU4Rec, GRU4Rec+, and NextItNet. Two possible reasons are: (1) these models are session-based methods without explicitly modeling the long-term user interests; (2) these methods equally treat all the items in a short context, which may fail to capture the short-term user intentions. Fifth, HGN outperforms BPRMF. Since BPRMF only captures the long-term interests of users, which does not incorporate the sequential patterns of user-item interactions. On the top of BPRMF, HGN adopts a hierarchical gating network to capture the sequential dynamics in the user actions and an item-item product module to explicitly capture the item-item relations, which leads to better performance.

\textbf{Other observations}. First, all the results reported on MovieLens-20M, GoodReads-Children and GoodReads-Comics are better than the results on other datasets, the major reason is that other datasets are more sparse and the data sparsity declines the recommendation performance. Second, SASRec outperforms Caser on most of the datasets. The main reason is that SASRec adaptively attends items that would reflect the short-term user interests. Third, SASRec and Caser achieve better performance than GRU4Rec, GRU4Rec+, and NextItNet in most cases. One possible reason is that SASRec and Caser both explicitly plug the user embeddings in their models, which allows the long-term user interests modeling. Fourth, GRU4Rec+ performs better than other methods on one dataset. The reason is that GRU4Rec+ not only captures the sequential patterns in the user-item sequence but also has a promising object function---\textit{BPR-max}. Fifth, all the methods perform better than BPR. This illustrates that only effectively modeling the long-term user interests is not sufficient to capture the user sequential behaviors.

\begin{table}[ht]
\centering
\caption{\label{tab:ablation_analysis}The ablation analysis on GoodReads-Comics and Amazon-Books datasets. \textit{F} denotes the feature gating module, \textit{I} denotes the instance gating module, \textit{avg} denotes the average pooling, and \textit{max} denotes the max pooling.}
\begin{tabular}{ |l|c|c|c|c| }
\hline
\multirow{2}{*}{Architecture} & \multicolumn{2}{c|}{\textit{Comics}} & \multicolumn{2}{c|}{\textit{Books}} \bigstrut \\\cline{2-5} 
& R@10 & N@10 & R@10 & N@10 \bigstrut \\ 
\hline
(1) BPR & 0.0911 & 0.0802 & 0.0310 & 0.0177 \\
(2) BPR+F+avg & 0.1555 & 0.1624 & 0.0361 & 0.0266 \\
(3) BPR+F+max & 0.1456 & 0.1550 & 0.0355 & 0.0240 \\
(4) BPR+I+avg & 0.1538 & 0.1591 & 0.0351 & 0.0254 \\
(5) BPR+I+max & 0.1489 & 0.1585 & 0.0329 & 0.0241 \\
(6) BPR+GRU & 0.1456 & 0.1581 & 0.0289 & 0.0216 \\
(7) BPR+CNN & 0.1305 & 0.1387 & 0.0278 & 0.0207 \\
(8) BPR+F+I+avg & 0.1635 & 0.1791 & 0.0391 & 0.0250 \\
(9) BPR+F+I+max & 0.1569 & 0.1658 & 0.0355 & 0.0234 \\
(10) HGN & \textbf{0.1743} & \textbf{0.1927} & \textbf{0.0429} & \textbf{0.0298} \\
\hline
\end{tabular}
\vspace{-0.3cm}
\end{table}

\subsection{Ablation Analysis} \label{sec:ablation}

To verify the effectiveness of the proposed feature gating, instance gating, and item-item product modules, we conduct an ablation analysis in Table \ref{tab:ablation_analysis} to demonstrate the importance each module contributes to the HGN model. In (1), we utilize only the BPR matrix factorization without any other components. In (2), we only incorporate the feature gating and apply the average pooling on the embeddings after the feature gating, on the top of (1). In (3), we replace the average pooling in (2) with max pooling. In (4), we only include the instance gating and apply the average pooling on the top of (1). In (5), we replace the average pooling in (4) with max-pooling. In (6), we adopt a recurrent neural network structure---gated recurrent unit (GRU) \cite{DBLP:conf/emnlp/ChoMGBBSB14} to learn the group-level representations of items. In (7), we replace the GRU in (6) with a convolutional neural network (CNN), where the structure and hyper-parameters are set the same in Caser \cite{DBLP:conf/wsdm/TangW18}. In (8), we both apply the feature and instance gating with average pooling. In (9), we replace the average pooling with max pooling. In (10), we present the overall HGN model to show the significance of the item-item product module.

From the results shown in Table \ref{tab:ablation_analysis}, we have some observations. First, from (1) and all others, we can observe that the conventional BPR matrix factorization to capture the long-term user interests cannot effectively model the short-term user interests. Second, from (2), (3), (4) and (5), the feature gating seems to achieve slightly better results than the instance gating. And the average pooling is slightly better than the max pooling, one possible reason is that the average pooling makes the representative item features accumulated, which results in a more effective representation of a group of $ |L| $ successive items. Third, from (6), (7), and (8), we observe that our hierarchical gating network achieves better performance than GRU and CNN but with \textit{fewer} learnable parameters\footnote{We verified the number of parameters of all three models by the \textit{named\_parameters()} function provided by PyTorch.} (if we set the item embedding size to 50 ($ d = 50 $), then the number of learnable parameters of our hierarchical gating network is 5,350, the number of parameters of the one-recurrent-layer GRU is 15,300, the number of parameters of the CNN in \cite{DBLP:conf/wsdm/TangW18} is 26,154). This result demonstrates that the proposed hierarchical gating network can effectively capture the sequential patterns in the user-item interaction sequence. Lastly, from (1), (8), and (9), we observe that by incorporating the item-item product, the performance further improves. The results demonstrate that explicitly capturing the relations between the items users accessed and those items users may interact in the future can provide a significant supplementary to model the user sequential dynamics.

\subsection{Training Efficiency} \label{sec:training_efficiency}
In this section, we evaluate the training efficiency with other state-of-the-art methods in terms of the training speed (time taken for one epoch of training). Since GRU4Rec+ has been compared with SASRec in \cite{DBLP:conf/icdm/KangM18}, we omit the training time comparison with GRU4Rec+. To make a fair comparison, we set the max sequence length of SASRec as $ 300 $ to cover more than 95\% of the sequence. All the experiments are conducted on a single GPU of Nvidia GeForce GTX 1080 Ti. All the compared methods are executed $ 20 $ epochs and we report the average computation time, which is shown in Table \ref{tab:training_time}. Note that the time reported only includes the training time of models without including the negative sampling time.

\begin{table}[ht]
\centering
\caption{\label{tab:training_time}The training time per epoch comparison on five datasets in terms of seconds.}
\begin{tabular}{ |c|c|c|c|c|c| }
 \hline
 & CDs & Books & ML20M & Children & Comics \\
\hline
HGN & \textbf{0.957s} & \textbf{2.086s} & \textbf{28.304s} & \textbf{3.496s} & \textbf{2.228s} \\
SASRec & 2.242s & 16.154s & 39.937s & 14.913s & 10.468s \\
Caser & 5.063s & 17.577s & 63.702s & 28.593s & 25.657s \\
\hline
\end{tabular}
\vspace{-0.3cm}
\end{table}

From the results in Table \ref{tab:training_time}, we can observe that HGN yields the fastest training speed on all datasets. As we have discussed in section \ref{sec:network training}, our model has less item complexity than SASRec, which is $ O(N^2d + Nd^2) $. Thus, our proposed model has better training efficiency both theoretically and practically.

\begin{table}[ht]
\centering
\caption{\label{tab:L_and_T}The effect of the length $ |L| $ and $ |T| $.}
\begin{tabular}{ |l|c|c|c|c| }
\hline
\multirow{2}{*}{Settings} & \multicolumn{2}{c|}{\textit{CDs}} & \multicolumn{2}{c|}{\textit{Comics}} \bigstrut \\\cline{2-5} 
& R@5 & R@10 & R@5 & R@10 \bigstrut \\ 
\hline
$|L|$=3, $|T|$=1 & 0.0260 & 0.0415 & 0.1202 & 0.1684 \\
$|L|$=3, $|T|$=2 & \textbf{0.0291} & 0.0448 & 0.1275 & 0.1758 \\
$|L|$=3, $|T|$=3 & 0.0289 & 0.0450 & \textbf{0.1296} & \textbf{0.1793} \\
$|L|$=5, $|T|$=1 & 0.0254 & 0.0417 & 0.1155 & 0.1645 \\
$|L|$=5, $|T|$=2 & 0.0261 & 0.0432 & 0.1215 & 0.1711 \\
$|L|$=5, $|T|$=3 & 0.0290 & \textbf{0.0456} & 0.1238 & 0.1738 \\
$|L|$=8, $|T|$=1 & 0.0220 & 0.0372 & 0.1083 & 0.1566 \\
$|L|$=8, $|T|$=2 & 0.0248 & 0.0401 & 0.1142 & 0.1636 \\
$|L|$=8, $|T|$=3 & 0.0260 & 0.0413 & 0.1160 & 0.1658 \\
\hline
\end{tabular}
\vspace{-0.3cm}
\end{table}

\subsection{The Sensitivity of Hyper-parameters} \label{subsec:parameter_sensitivity}

\begin{figure}[t!]
    \centering
    \begin{subfigure}[t]{0.25\textwidth}
        \centering
        \includegraphics[width=\linewidth]{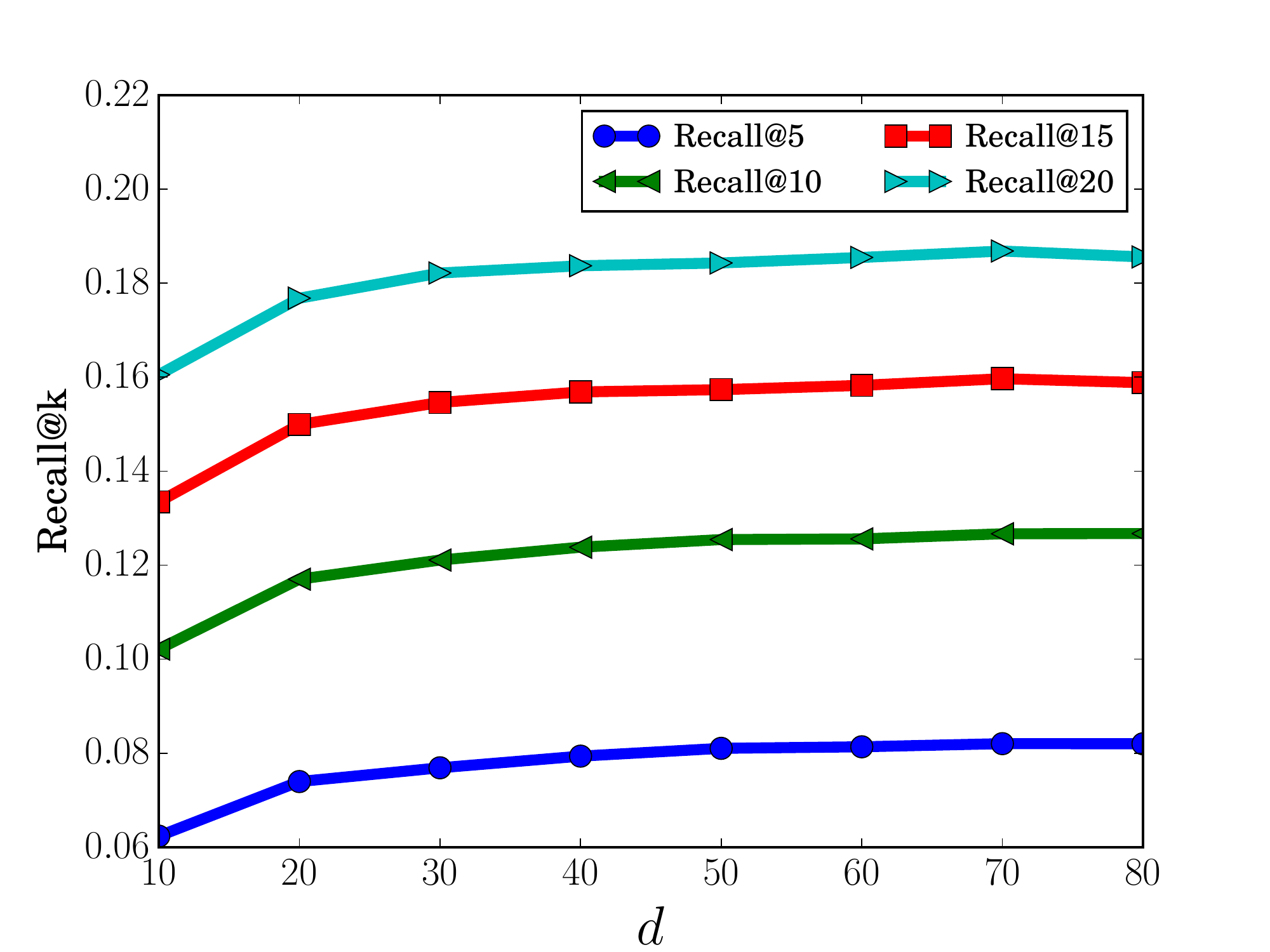}
        \caption{\label{fig:CDs_da_var}$ d $ on Children}
    \end{subfigure}%
    \begin{subfigure}[t]{0.25\textwidth}
        \centering
        \includegraphics[width=\linewidth]{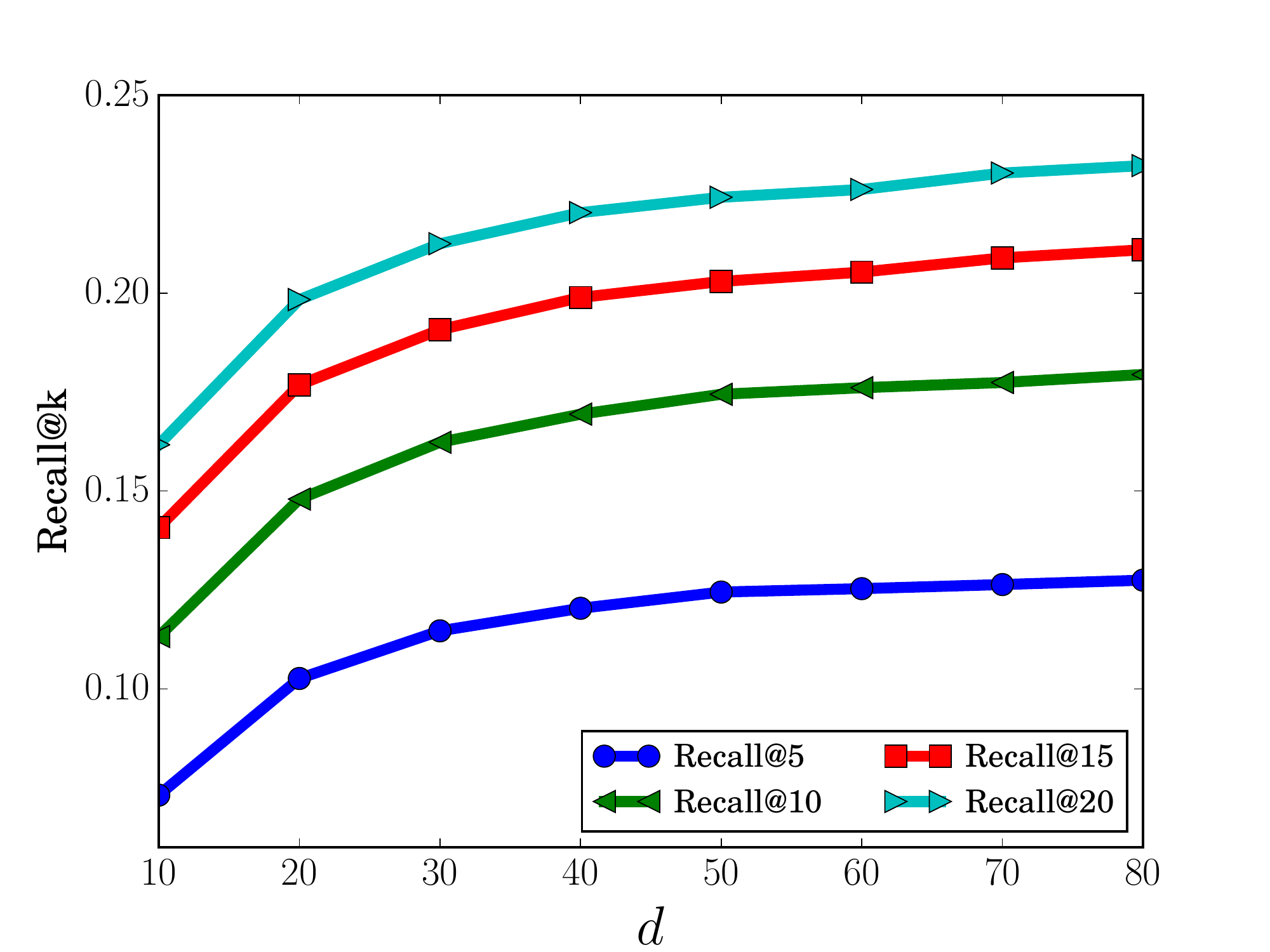}
        \caption{\label{fig:Foursquare_da_var}$ d $ on Comics}
    \end{subfigure}
    \caption{\label{fig:d_var}The dimension variations of embeddings.}
\vspace{-0.3cm}
\end{figure}

We present the effect of two hyper-parameters: the dimension of the item embeddings $ d $ and the length of successive items $ |L| $ and $ |T| $. The effects of these two parameters are shown in Figure \ref{fig:d_var} and Table \ref{tab:L_and_T}. Due to the space limit, we only present the effects on two datasets, the parameter effects on other datasets have similar trends.

The variation of $ d $ is shown in Figure \ref{fig:d_var}. We can observe that a small dimension of item embeddings is not sufficient to express the latent features of items. By increasing the dimension of item embeddings, the model has more capacity to model the complex features of items. With the increase of $ d $, the model performance largely improves and becomes steady.

The variation of $ |L|$ and $ |T| $ is shown in Table \ref{tab:L_and_T}. We observe that when $ |L| $ is fixed, a larger value of $ |T| $, i.e. 3, can achieve better performance. This may illustrate that a group of $ |L| $ items may determine several items that user will interact in the near future. We also observe that smaller $ |L| $ has better results than larger ones. One possible reason is that larger $ |L| $ may include too many irrelevant items for predicting future items.

\section{Conclusion}
In this paper, we propose a hierarchical gating network with an item-item product module for the sequential recommendation. The model adopts a feature gating module and an instance gating module to control what item features can be passed to downstream layers, where informative latent features and items can be selected. Moreover, we apply an item-item product module to capture the relations between closely relevant items. Experimental results on five real-world datasets clearly validate the performance of our model over many state-of-the-art methods and demonstrate the effectiveness of the gating and item-item product modules.

\bibliographystyle{ACM-Reference-Format}
\bibliography{seq} 


\begin{thebibliography}{42}


\ifx \showCODEN    \undefined \def \showCODEN     #1{\unskip}     \fi
\ifx \showDOI      \undefined \def \showDOI       #1{#1}\fi
\ifx \showISBNx    \undefined \def \showISBNx     #1{\unskip}     \fi
\ifx \showISBNxiii \undefined \def \showISBNxiii  #1{\unskip}     \fi
\ifx \showISSN     \undefined \def \showISSN      #1{\unskip}     \fi
\ifx \showLCCN     \undefined \def \showLCCN      #1{\unskip}     \fi
\ifx \shownote     \undefined \def \shownote      #1{#1}          \fi
\ifx \showarticletitle \undefined \def \showarticletitle #1{#1}   \fi
\ifx \showURL      \undefined \def \showURL       {\relax}        \fi
\providecommand\bibfield[2]{#2}
\providecommand\bibinfo[2]{#2}
\providecommand\natexlab[1]{#1}
\providecommand\showeprint[2][]{arXiv:#2}

\bibitem[\protect\citeauthoryear{Chen, Xu, Zhang, Tang, Cao, Qin, and Zha}{Chen
  et~al\mbox{.}}{2018}]%
        {DBLP:conf/wsdm/ChenXZT0QZ18}
\bibfield{author}{\bibinfo{person}{Xu Chen}, \bibinfo{person}{Hongteng Xu},
  \bibinfo{person}{Yongfeng Zhang}, \bibinfo{person}{Jiaxi Tang},
  \bibinfo{person}{Yixin Cao}, \bibinfo{person}{Zheng Qin}, {and}
  \bibinfo{person}{Hongyuan Zha}.} \bibinfo{year}{2018}\natexlab{}.
\newblock \showarticletitle{Sequential Recommendation with User Memory
  Networks}. In \bibinfo{booktitle}{\emph{{WSDM}}}. \bibinfo{publisher}{{ACM}},
  \bibinfo{pages}{108--116}.
\newblock


\bibitem[\protect\citeauthoryear{Cheng, Yang, Lyu, and King}{Cheng
  et~al\mbox{.}}{2013}]%
        {DBLP:conf/ijcai/ChengYLK13}
\bibfield{author}{\bibinfo{person}{Chen Cheng}, \bibinfo{person}{Haiqin Yang},
  \bibinfo{person}{Michael~R. Lyu}, {and} \bibinfo{person}{Irwin King}.}
  \bibinfo{year}{2013}\natexlab{}.
\newblock \showarticletitle{Where You Like to Go Next: Successive
  Point-of-Interest Recommendation}. In \bibinfo{booktitle}{\emph{{IJCAI}}}.
  \bibinfo{publisher}{{IJCAI/AAAI}}, \bibinfo{pages}{2605--2611}.
\newblock


\bibitem[\protect\citeauthoryear{Cho, van Merrienboer, G{\"{u}}l{\c{c}}ehre,
  Bahdanau, Bougares, Schwenk, and Bengio}{Cho et~al\mbox{.}}{2014}]%
        {DBLP:conf/emnlp/ChoMGBBSB14}
\bibfield{author}{\bibinfo{person}{Kyunghyun Cho}, \bibinfo{person}{Bart van
  Merrienboer}, \bibinfo{person}{{\c{C}}aglar G{\"{u}}l{\c{c}}ehre},
  \bibinfo{person}{Dzmitry Bahdanau}, \bibinfo{person}{Fethi Bougares},
  \bibinfo{person}{Holger Schwenk}, {and} \bibinfo{person}{Yoshua Bengio}.}
  \bibinfo{year}{2014}\natexlab{}.
\newblock \showarticletitle{Learning Phrase Representations using {RNN}
  Encoder-Decoder for Statistical Machine Translation}. In
  \bibinfo{booktitle}{\emph{{EMNLP}}}. \bibinfo{publisher}{{ACL}},
  \bibinfo{pages}{1724--1734}.
\newblock


\bibitem[\protect\citeauthoryear{Dauphin, Fan, Auli, and Grangier}{Dauphin
  et~al\mbox{.}}{2017}]%
        {DBLP:conf/icml/DauphinFAG17}
\bibfield{author}{\bibinfo{person}{Yann~N. Dauphin}, \bibinfo{person}{Angela
  Fan}, \bibinfo{person}{Michael Auli}, {and} \bibinfo{person}{David
  Grangier}.} \bibinfo{year}{2017}\natexlab{}.
\newblock \showarticletitle{Language Modeling with Gated Convolutional
  Networks}. In \bibinfo{booktitle}{\emph{{ICML}}}
  \emph{(\bibinfo{series}{Proceedings of Machine Learning Research})},
  Vol.~\bibinfo{volume}{70}. \bibinfo{publisher}{{PMLR}},
  \bibinfo{pages}{933--941}.
\newblock


\bibitem[\protect\citeauthoryear{Guo, Tang, Ye, Li, and He}{Guo
  et~al\mbox{.}}{2017}]%
        {DBLP:conf/ijcai/GuoTYLH17}
\bibfield{author}{\bibinfo{person}{Huifeng Guo}, \bibinfo{person}{Ruiming
  Tang}, \bibinfo{person}{Yunming Ye}, \bibinfo{person}{Zhenguo Li}, {and}
  \bibinfo{person}{Xiuqiang He}.} \bibinfo{year}{2017}\natexlab{}.
\newblock \showarticletitle{DeepFM: {A} Factorization-Machine based Neural
  Network for {CTR} Prediction}. In \bibinfo{booktitle}{\emph{{IJCAI}}}.
  \bibinfo{publisher}{ijcai.org}, \bibinfo{pages}{1725--1731}.
\newblock


\bibitem[\protect\citeauthoryear{Harper and Konstan}{Harper and
  Konstan}{2016}]%
        {DBLP:journals/tiis/HarperK16}
\bibfield{author}{\bibinfo{person}{F.~Maxwell Harper} {and}
  \bibinfo{person}{Joseph~A. Konstan}.} \bibinfo{year}{2016}\natexlab{}.
\newblock \showarticletitle{The MovieLens Datasets: History and Context}.
\newblock \bibinfo{journal}{\emph{TiiS}} \bibinfo{volume}{5},
  \bibinfo{number}{4} (\bibinfo{year}{2016}), \bibinfo{pages}{19:1--19:19}.
\newblock


\bibitem[\protect\citeauthoryear{He, Kang, and McAuley}{He
  et~al\mbox{.}}{2017a}]%
        {DBLP:conf/recsys/HeKM17}
\bibfield{author}{\bibinfo{person}{Ruining He}, \bibinfo{person}{Wang{-}Cheng
  Kang}, {and} \bibinfo{person}{Julian McAuley}.}
  \bibinfo{year}{2017}\natexlab{a}.
\newblock \showarticletitle{Translation-based Recommendation}. In
  \bibinfo{booktitle}{\emph{RecSys}}. \bibinfo{publisher}{{ACM}},
  \bibinfo{pages}{161--169}.
\newblock


\bibitem[\protect\citeauthoryear{He and McAuley}{He and McAuley}{2016a}]%
        {DBLP:conf/icdm/HeM16}
\bibfield{author}{\bibinfo{person}{Ruining He} {and} \bibinfo{person}{Julian
  McAuley}.} \bibinfo{year}{2016}\natexlab{a}.
\newblock \showarticletitle{Fusing Similarity Models with Markov Chains for
  Sparse Sequential Recommendation}. In \bibinfo{booktitle}{\emph{{ICDM}}}.
  \bibinfo{publisher}{{IEEE}}, \bibinfo{pages}{191--200}.
\newblock


\bibitem[\protect\citeauthoryear{He and McAuley}{He and McAuley}{2016b}]%
        {DBLP:conf/www/HeM16}
\bibfield{author}{\bibinfo{person}{Ruining He} {and} \bibinfo{person}{Julian
  McAuley}.} \bibinfo{year}{2016}\natexlab{b}.
\newblock \showarticletitle{Ups and Downs: Modeling the Visual Evolution of
  Fashion Trends with One-Class Collaborative Filtering}. In
  \bibinfo{booktitle}{\emph{{WWW}}}. \bibinfo{publisher}{{ACM}},
  \bibinfo{pages}{507--517}.
\newblock


\bibitem[\protect\citeauthoryear{He, Liao, Zhang, Nie, Hu, and Chua}{He
  et~al\mbox{.}}{2017b}]%
        {DBLP:conf/www/HeLZNHC17}
\bibfield{author}{\bibinfo{person}{Xiangnan He}, \bibinfo{person}{Lizi Liao},
  \bibinfo{person}{Hanwang Zhang}, \bibinfo{person}{Liqiang Nie},
  \bibinfo{person}{Xia Hu}, {and} \bibinfo{person}{Tat{-}Seng Chua}.}
  \bibinfo{year}{2017}\natexlab{b}.
\newblock \showarticletitle{Neural Collaborative Filtering}. In
  \bibinfo{booktitle}{\emph{{WWW}}}. \bibinfo{publisher}{{ACM}},
  \bibinfo{pages}{173--182}.
\newblock


\bibitem[\protect\citeauthoryear{He, Zhang, Kan, and Chua}{He
  et~al\mbox{.}}{2016}]%
        {DBLP:conf/sigir/HeZKC16}
\bibfield{author}{\bibinfo{person}{Xiangnan He}, \bibinfo{person}{Hanwang
  Zhang}, \bibinfo{person}{Min{-}Yen Kan}, {and} \bibinfo{person}{Tat{-}Seng
  Chua}.} \bibinfo{year}{2016}\natexlab{}.
\newblock \showarticletitle{Fast Matrix Factorization for Online Recommendation
  with Implicit Feedback}. In \bibinfo{booktitle}{\emph{{SIGIR}}}.
  \bibinfo{publisher}{{ACM}}, \bibinfo{pages}{549--558}.
\newblock


\bibitem[\protect\citeauthoryear{Hidasi and Karatzoglou}{Hidasi and
  Karatzoglou}{2018}]%
        {DBLP:conf/cikm/HidasiK18}
\bibfield{author}{\bibinfo{person}{Bal{\'{a}}zs Hidasi} {and}
  \bibinfo{person}{Alexandros Karatzoglou}.} \bibinfo{year}{2018}\natexlab{}.
\newblock \showarticletitle{Recurrent Neural Networks with Top-k Gains for
  Session-based Recommendations}. In \bibinfo{booktitle}{\emph{{CIKM}}}.
  \bibinfo{publisher}{{ACM}}, \bibinfo{pages}{843--852}.
\newblock


\bibitem[\protect\citeauthoryear{Hidasi, Karatzoglou, Baltrunas, and
  Tikk}{Hidasi et~al\mbox{.}}{2015}]%
        {DBLP:journals/corr/HidasiKBT15}
\bibfield{author}{\bibinfo{person}{Bal{\'{a}}zs Hidasi},
  \bibinfo{person}{Alexandros Karatzoglou}, \bibinfo{person}{Linas Baltrunas},
  {and} \bibinfo{person}{Domonkos Tikk}.} \bibinfo{year}{2015}\natexlab{}.
\newblock \showarticletitle{Session-based Recommendations with Recurrent Neural
  Networks}.
\newblock \bibinfo{journal}{\emph{CoRR}}  \bibinfo{volume}{abs/1511.06939}
  (\bibinfo{year}{2015}).
\newblock


\bibitem[\protect\citeauthoryear{Hsieh, Yang, Cui, Lin, Belongie, and
  Estrin}{Hsieh et~al\mbox{.}}{2017}]%
        {DBLP:conf/www/HsiehYCLBE17}
\bibfield{author}{\bibinfo{person}{Cheng{-}Kang Hsieh}, \bibinfo{person}{Longqi
  Yang}, \bibinfo{person}{Yin Cui}, \bibinfo{person}{Tsung{-}Yi Lin},
  \bibinfo{person}{Serge~J. Belongie}, {and} \bibinfo{person}{Deborah Estrin}.}
  \bibinfo{year}{2017}\natexlab{}.
\newblock \showarticletitle{Collaborative Metric Learning}. In
  \bibinfo{booktitle}{\emph{{WWW}}}. \bibinfo{publisher}{{ACM}},
  \bibinfo{pages}{193--201}.
\newblock


\bibitem[\protect\citeauthoryear{Hu, Koren, and Volinsky}{Hu
  et~al\mbox{.}}{2008}]%
        {DBLP:conf/icdm/HuKV08}
\bibfield{author}{\bibinfo{person}{Yifan Hu}, \bibinfo{person}{Yehuda Koren},
  {and} \bibinfo{person}{Chris Volinsky}.} \bibinfo{year}{2008}\natexlab{}.
\newblock \showarticletitle{Collaborative Filtering for Implicit Feedback
  Datasets}. In \bibinfo{booktitle}{\emph{{ICDM}}}. \bibinfo{publisher}{{IEEE}
  Computer Society}, \bibinfo{pages}{263--272}.
\newblock


\bibitem[\protect\citeauthoryear{Huang, Zhao, Dou, Wen, and Chang}{Huang
  et~al\mbox{.}}{2018}]%
        {DBLP:conf/sigir/HuangZDWC18}
\bibfield{author}{\bibinfo{person}{Jin Huang}, \bibinfo{person}{Wayne~Xin
  Zhao}, \bibinfo{person}{Hong{-}Jian Dou}, \bibinfo{person}{Ji{-}Rong Wen},
  {and} \bibinfo{person}{Edward~Y. Chang}.} \bibinfo{year}{2018}\natexlab{}.
\newblock \showarticletitle{Improving Sequential Recommendation with
  Knowledge-Enhanced Memory Networks}. In \bibinfo{booktitle}{\emph{{SIGIR}}}.
  \bibinfo{publisher}{{ACM}}, \bibinfo{pages}{505--514}.
\newblock


\bibitem[\protect\citeauthoryear{Kabbur, Ning, and Karypis}{Kabbur
  et~al\mbox{.}}{2013}]%
        {DBLP:conf/kdd/KabburNK13}
\bibfield{author}{\bibinfo{person}{Santosh Kabbur}, \bibinfo{person}{Xia Ning},
  {and} \bibinfo{person}{George Karypis}.} \bibinfo{year}{2013}\natexlab{}.
\newblock \showarticletitle{{FISM:} factored item similarity models for top-N
  recommender systems}. In \bibinfo{booktitle}{\emph{{KDD}}}.
  \bibinfo{publisher}{{ACM}}, \bibinfo{pages}{659--667}.
\newblock


\bibitem[\protect\citeauthoryear{Kang and McAuley}{Kang and McAuley}{2018}]%
        {DBLP:conf/icdm/KangM18}
\bibfield{author}{\bibinfo{person}{Wang{-}Cheng Kang} {and}
  \bibinfo{person}{Julian McAuley}.} \bibinfo{year}{2018}\natexlab{}.
\newblock \showarticletitle{Self-Attentive Sequential Recommendation}. In
  \bibinfo{booktitle}{\emph{{ICDM}}}. \bibinfo{publisher}{{IEEE} Computer
  Society}, \bibinfo{pages}{197--206}.
\newblock


\bibitem[\protect\citeauthoryear{Kingma and Ba}{Kingma and Ba}{2014}]%
        {DBLP:journals/corr/KingmaB14}
\bibfield{author}{\bibinfo{person}{Diederik~P. Kingma} {and}
  \bibinfo{person}{Jimmy Ba}.} \bibinfo{year}{2014}\natexlab{}.
\newblock \showarticletitle{Adam: {A} Method for Stochastic Optimization}.
\newblock \bibinfo{journal}{\emph{CoRR}}  \bibinfo{volume}{abs/1412.6980}
  (\bibinfo{year}{2014}).
\newblock


\bibitem[\protect\citeauthoryear{Li, Ren, Chen, Ren, Lian, and Ma}{Li
  et~al\mbox{.}}{2017}]%
        {DBLP:conf/cikm/LiRCRLM17}
\bibfield{author}{\bibinfo{person}{Jing Li}, \bibinfo{person}{Pengjie Ren},
  \bibinfo{person}{Zhumin Chen}, \bibinfo{person}{Zhaochun Ren},
  \bibinfo{person}{Tao Lian}, {and} \bibinfo{person}{Jun Ma}.}
  \bibinfo{year}{2017}\natexlab{}.
\newblock \showarticletitle{Neural Attentive Session-based Recommendation}. In
  \bibinfo{booktitle}{\emph{{CIKM}}}. \bibinfo{publisher}{{ACM}},
  \bibinfo{pages}{1419--1428}.
\newblock


\bibitem[\protect\citeauthoryear{Lian, Zhou, Zhang, Chen, Xie, and Sun}{Lian
  et~al\mbox{.}}{2018}]%
        {DBLP:conf/kdd/LianZZCXS18}
\bibfield{author}{\bibinfo{person}{Jianxun Lian}, \bibinfo{person}{Xiaohuan
  Zhou}, \bibinfo{person}{Fuzheng Zhang}, \bibinfo{person}{Zhongxia Chen},
  \bibinfo{person}{Xing Xie}, {and} \bibinfo{person}{Guangzhong Sun}.}
  \bibinfo{year}{2018}\natexlab{}.
\newblock \showarticletitle{xDeepFM: Combining Explicit and Implicit Feature
  Interactions for Recommender Systems}. In \bibinfo{booktitle}{\emph{{KDD}}}.
  \bibinfo{publisher}{{ACM}}, \bibinfo{pages}{1754--1763}.
\newblock


\bibitem[\protect\citeauthoryear{Liang, Charlin, McInerney, and Blei}{Liang
  et~al\mbox{.}}{2016}]%
        {DBLP:conf/www/LiangCMB16}
\bibfield{author}{\bibinfo{person}{Dawen Liang}, \bibinfo{person}{Laurent
  Charlin}, \bibinfo{person}{James McInerney}, {and} \bibinfo{person}{David~M.
  Blei}.} \bibinfo{year}{2016}\natexlab{}.
\newblock \showarticletitle{Modeling User Exposure in Recommendation}. In
  \bibinfo{booktitle}{\emph{{WWW}}}. \bibinfo{publisher}{{ACM}},
  \bibinfo{pages}{951--961}.
\newblock


\bibitem[\protect\citeauthoryear{Ma, Kang, Wu, Wang, and Liu}{Ma
  et~al\mbox{.}}{2019}]%
        {DBLP:conf/wsdm/MaKWWL19}
\bibfield{author}{\bibinfo{person}{Chen Ma}, \bibinfo{person}{Peng Kang},
  \bibinfo{person}{Bin Wu}, \bibinfo{person}{Qinglong Wang}, {and}
  \bibinfo{person}{Xue Liu}.} \bibinfo{year}{2019}\natexlab{}.
\newblock \showarticletitle{Gated Attentive-Autoencoder for Content-Aware
  Recommendation}. In \bibinfo{booktitle}{\emph{{WSDM}}}.
  \bibinfo{publisher}{{ACM}}, \bibinfo{pages}{519--527}.
\newblock


\bibitem[\protect\citeauthoryear{Ma, Zhang, Wang, and Liu}{Ma
  et~al\mbox{.}}{2018}]%
        {DBLP:conf/cikm/MaZWL18}
\bibfield{author}{\bibinfo{person}{Chen Ma}, \bibinfo{person}{Yingxue Zhang},
  \bibinfo{person}{Qinglong Wang}, {and} \bibinfo{person}{Xue Liu}.}
  \bibinfo{year}{2018}\natexlab{}.
\newblock \showarticletitle{Point-of-Interest Recommendation: Exploiting
  Self-Attentive Autoencoders with Neighbor-Aware Influence}. In
  \bibinfo{booktitle}{\emph{{CIKM}}}. \bibinfo{publisher}{{ACM}},
  \bibinfo{pages}{697--706}.
\newblock


\bibitem[\protect\citeauthoryear{Ning, Desrosiers, and Karypis}{Ning
  et~al\mbox{.}}{2015}]%
        {DBLP:reference/sp/NingDK15}
\bibfield{author}{\bibinfo{person}{Xia Ning}, \bibinfo{person}{Christian
  Desrosiers}, {and} \bibinfo{person}{George Karypis}.}
  \bibinfo{year}{2015}\natexlab{}.
\newblock \showarticletitle{A Comprehensive Survey of Neighborhood-Based
  Recommendation Methods}.
\newblock In \bibinfo{booktitle}{\emph{Recommender Systems Handbook}}.
  \bibinfo{publisher}{Springer}, \bibinfo{pages}{37--76}.
\newblock


\bibitem[\protect\citeauthoryear{Pan, Zhou, Cao, Liu, Lukose, Scholz, and
  Yang}{Pan et~al\mbox{.}}{2008}]%
        {DBLP:conf/icdm/PanZCLLSY08}
\bibfield{author}{\bibinfo{person}{Rong Pan}, \bibinfo{person}{Yunhong Zhou},
  \bibinfo{person}{Bin Cao}, \bibinfo{person}{Nathan~Nan Liu},
  \bibinfo{person}{Rajan~M. Lukose}, \bibinfo{person}{Martin Scholz}, {and}
  \bibinfo{person}{Qiang Yang}.} \bibinfo{year}{2008}\natexlab{}.
\newblock \showarticletitle{One-Class Collaborative Filtering}. In
  \bibinfo{booktitle}{\emph{{ICDM}}}. \bibinfo{publisher}{{IEEE} Computer
  Society}, \bibinfo{pages}{502--511}.
\newblock


\bibitem[\protect\citeauthoryear{Pei, Yang, Sun, Zhang, Bozzon, and Tax}{Pei
  et~al\mbox{.}}{2017}]%
        {DBLP:conf/cikm/PeiYSZBT17}
\bibfield{author}{\bibinfo{person}{Wenjie Pei}, \bibinfo{person}{Jie Yang},
  \bibinfo{person}{Zhu Sun}, \bibinfo{person}{Jie Zhang},
  \bibinfo{person}{Alessandro Bozzon}, {and} \bibinfo{person}{David M.~J.
  Tax}.} \bibinfo{year}{2017}\natexlab{}.
\newblock \showarticletitle{Interacting Attention-gated Recurrent Networks for
  Recommendation}. In \bibinfo{booktitle}{\emph{{CIKM}}}.
  \bibinfo{publisher}{{ACM}}, \bibinfo{pages}{1459--1468}.
\newblock


\bibitem[\protect\citeauthoryear{Quadrana, Karatzoglou, Hidasi, and
  Cremonesi}{Quadrana et~al\mbox{.}}{2017}]%
        {DBLP:conf/recsys/QuadranaKHC17}
\bibfield{author}{\bibinfo{person}{Massimo Quadrana},
  \bibinfo{person}{Alexandros Karatzoglou}, \bibinfo{person}{Bal{\'{a}}zs
  Hidasi}, {and} \bibinfo{person}{Paolo Cremonesi}.}
  \bibinfo{year}{2017}\natexlab{}.
\newblock \showarticletitle{Personalizing Session-based Recommendations with
  Hierarchical Recurrent Neural Networks}. In
  \bibinfo{booktitle}{\emph{RecSys}}. \bibinfo{publisher}{{ACM}},
  \bibinfo{pages}{130--137}.
\newblock


\bibitem[\protect\citeauthoryear{Rendle, Freudenthaler, Gantner, and
  Schmidt{-}Thieme}{Rendle et~al\mbox{.}}{2009}]%
        {DBLP:conf/uai/RendleFGS09}
\bibfield{author}{\bibinfo{person}{Steffen Rendle}, \bibinfo{person}{Christoph
  Freudenthaler}, \bibinfo{person}{Zeno Gantner}, {and} \bibinfo{person}{Lars
  Schmidt{-}Thieme}.} \bibinfo{year}{2009}\natexlab{}.
\newblock \showarticletitle{{BPR:} Bayesian Personalized Ranking from Implicit
  Feedback}. In \bibinfo{booktitle}{\emph{{UAI}}}. \bibinfo{publisher}{{AUAI}
  Press}, \bibinfo{pages}{452--461}.
\newblock


\bibitem[\protect\citeauthoryear{Rendle, Freudenthaler, and
  Schmidt{-}Thieme}{Rendle et~al\mbox{.}}{2010}]%
        {DBLP:conf/www/RendleFS10}
\bibfield{author}{\bibinfo{person}{Steffen Rendle}, \bibinfo{person}{Christoph
  Freudenthaler}, {and} \bibinfo{person}{Lars Schmidt{-}Thieme}.}
  \bibinfo{year}{2010}\natexlab{}.
\newblock \showarticletitle{Factorizing personalized Markov chains for
  next-basket recommendation}. In \bibinfo{booktitle}{\emph{{WWW}}}.
  \bibinfo{publisher}{{ACM}}, \bibinfo{pages}{811--820}.
\newblock


\bibitem[\protect\citeauthoryear{Salakhutdinov, Mnih, and Hinton}{Salakhutdinov
  et~al\mbox{.}}{2007}]%
        {DBLP:conf/icml/SalakhutdinovMH07}
\bibfield{author}{\bibinfo{person}{Ruslan Salakhutdinov},
  \bibinfo{person}{Andriy Mnih}, {and} \bibinfo{person}{Geoffrey~E. Hinton}.}
  \bibinfo{year}{2007}\natexlab{}.
\newblock \showarticletitle{Restricted Boltzmann machines for collaborative
  filtering}. In \bibinfo{booktitle}{\emph{{ICML}}}
  \emph{(\bibinfo{series}{{ACM} International Conference Proceeding Series})},
  Vol.~\bibinfo{volume}{227}. \bibinfo{publisher}{{ACM}},
  \bibinfo{pages}{791--798}.
\newblock


\bibitem[\protect\citeauthoryear{Sarwar, Karypis, Konstan, and Riedl}{Sarwar
  et~al\mbox{.}}{2001}]%
        {DBLP:conf/www/SarwarKKR01}
\bibfield{author}{\bibinfo{person}{Badrul~Munir Sarwar},
  \bibinfo{person}{George Karypis}, \bibinfo{person}{Joseph~A. Konstan}, {and}
  \bibinfo{person}{John Riedl}.} \bibinfo{year}{2001}\natexlab{}.
\newblock \showarticletitle{Item-based collaborative filtering recommendation
  algorithms}. In \bibinfo{booktitle}{\emph{{WWW}}}.
  \bibinfo{publisher}{{ACM}}, \bibinfo{pages}{285--295}.
\newblock


\bibitem[\protect\citeauthoryear{Sukhbaatar, Szlam, Weston, and
  Fergus}{Sukhbaatar et~al\mbox{.}}{2015}]%
        {DBLP:conf/nips/SukhbaatarSWF15}
\bibfield{author}{\bibinfo{person}{Sainbayar Sukhbaatar},
  \bibinfo{person}{Arthur Szlam}, \bibinfo{person}{Jason Weston}, {and}
  \bibinfo{person}{Rob Fergus}.} \bibinfo{year}{2015}\natexlab{}.
\newblock \showarticletitle{End-To-End Memory Networks}. In
  \bibinfo{booktitle}{\emph{{NIPS}}}. \bibinfo{pages}{2440--2448}.
\newblock


\bibitem[\protect\citeauthoryear{Tang and Wang}{Tang and Wang}{2018}]%
        {DBLP:conf/wsdm/TangW18}
\bibfield{author}{\bibinfo{person}{Jiaxi Tang} {and} \bibinfo{person}{Ke
  Wang}.} \bibinfo{year}{2018}\natexlab{}.
\newblock \showarticletitle{Personalized Top-N Sequential Recommendation via
  Convolutional Sequence Embedding}. In \bibinfo{booktitle}{\emph{{WSDM}}}.
  \bibinfo{publisher}{{ACM}}, \bibinfo{pages}{565--573}.
\newblock


\bibitem[\protect\citeauthoryear{Tran, Lee, Liao, and Lee}{Tran
  et~al\mbox{.}}{2018}]%
        {DBLP:conf/cikm/TranLL018}
\bibfield{author}{\bibinfo{person}{Thanh Tran}, \bibinfo{person}{Kyumin Lee},
  \bibinfo{person}{Yiming Liao}, {and} \bibinfo{person}{Dongwon Lee}.}
  \bibinfo{year}{2018}\natexlab{}.
\newblock \showarticletitle{Regularizing Matrix Factorization with User and
  Item Embeddings for Recommendation}. In \bibinfo{booktitle}{\emph{{CIKM}}}.
  \bibinfo{publisher}{{ACM}}, \bibinfo{pages}{687--696}.
\newblock


\bibitem[\protect\citeauthoryear{Tran, Liu, Lee, and Kong}{Tran
  et~al\mbox{.}}{[n.d.]}]%
        {DBLP:conf/www/TranLLK19}
\bibfield{author}{\bibinfo{person}{Thanh Tran}, \bibinfo{person}{Xinyue Liu},
  \bibinfo{person}{Kyumin Lee}, {and} \bibinfo{person}{Xiangnan Kong}.}
  \bibinfo{year}{[n.d.]}\natexlab{}.
\newblock \showarticletitle{Signed Distance-based Deep Memory Recommender}.
\newblock


\bibitem[\protect\citeauthoryear{Vaswani, Shazeer, Parmar, Uszkoreit, Jones,
  Gomez, Kaiser, and Polosukhin}{Vaswani et~al\mbox{.}}{2017}]%
        {DBLP:conf/nips/VaswaniSPUJGKP17}
\bibfield{author}{\bibinfo{person}{Ashish Vaswani}, \bibinfo{person}{Noam
  Shazeer}, \bibinfo{person}{Niki Parmar}, \bibinfo{person}{Jakob Uszkoreit},
  \bibinfo{person}{Llion Jones}, \bibinfo{person}{Aidan~N. Gomez},
  \bibinfo{person}{Lukasz Kaiser}, {and} \bibinfo{person}{Illia Polosukhin}.}
  \bibinfo{year}{2017}\natexlab{}.
\newblock \showarticletitle{Attention is All you Need}. In
  \bibinfo{booktitle}{\emph{{NIPS}}}. \bibinfo{pages}{6000--6010}.
\newblock


\bibitem[\protect\citeauthoryear{Wan and McAuley}{Wan and McAuley}{2018}]%
        {DBLP:conf/recsys/WanM18}
\bibfield{author}{\bibinfo{person}{Mengting Wan} {and} \bibinfo{person}{Julian
  McAuley}.} \bibinfo{year}{2018}\natexlab{}.
\newblock \showarticletitle{Item recommendation on monotonic behavior chains}.
  In \bibinfo{booktitle}{\emph{RecSys}}. \bibinfo{publisher}{{ACM}},
  \bibinfo{pages}{86--94}.
\newblock


\bibitem[\protect\citeauthoryear{Wang, Wang, and Yeung}{Wang
  et~al\mbox{.}}{2015}]%
        {DBLP:conf/kdd/WangWY15}
\bibfield{author}{\bibinfo{person}{Hao Wang}, \bibinfo{person}{Naiyan Wang},
  {and} \bibinfo{person}{Dit{-}Yan Yeung}.} \bibinfo{year}{2015}\natexlab{}.
\newblock \showarticletitle{Collaborative Deep Learning for Recommender
  Systems}. In \bibinfo{booktitle}{\emph{{KDD}}}. \bibinfo{publisher}{{ACM}},
  \bibinfo{pages}{1235--1244}.
\newblock


\bibitem[\protect\citeauthoryear{Wu, DuBois, Zheng, and Ester}{Wu
  et~al\mbox{.}}{2016}]%
        {DBLP:conf/wsdm/WuDZE16}
\bibfield{author}{\bibinfo{person}{Yao Wu}, \bibinfo{person}{Christopher
  DuBois}, \bibinfo{person}{Alice~X. Zheng}, {and} \bibinfo{person}{Martin
  Ester}.} \bibinfo{year}{2016}\natexlab{}.
\newblock \showarticletitle{Collaborative Denoising Auto-Encoders for Top-N
  Recommender Systems}. In \bibinfo{booktitle}{\emph{{WSDM}}}.
  \bibinfo{publisher}{{ACM}}, \bibinfo{pages}{153--162}.
\newblock


\bibitem[\protect\citeauthoryear{Xue, Dai, Zhang, Huang, and Chen}{Xue
  et~al\mbox{.}}{2017}]%
        {DBLP:conf/ijcai/XueDZHC17}
\bibfield{author}{\bibinfo{person}{Hong{-}Jian Xue}, \bibinfo{person}{Xinyu
  Dai}, \bibinfo{person}{Jianbing Zhang}, \bibinfo{person}{Shujian Huang},
  {and} \bibinfo{person}{Jiajun Chen}.} \bibinfo{year}{2017}\natexlab{}.
\newblock \showarticletitle{Deep Matrix Factorization Models for Recommender
  Systems}. In \bibinfo{booktitle}{\emph{{IJCAI}}}.
  \bibinfo{publisher}{ijcai.org}, \bibinfo{pages}{3203--3209}.
\newblock


\bibitem[\protect\citeauthoryear{Yuan, Karatzoglou, Arapakis, Jose, and
  He}{Yuan et~al\mbox{.}}{2019}]%
        {DBLP:conf/wsdm/YuanKAJ019}
\bibfield{author}{\bibinfo{person}{Fajie Yuan}, \bibinfo{person}{Alexandros
  Karatzoglou}, \bibinfo{person}{Ioannis Arapakis}, \bibinfo{person}{Joemon~M.
  Jose}, {and} \bibinfo{person}{Xiangnan He}.} \bibinfo{year}{2019}\natexlab{}.
\newblock \showarticletitle{A Simple Convolutional Generative Network for Next
  Item Recommendation}. In \bibinfo{booktitle}{\emph{{WSDM}}}.
  \bibinfo{publisher}{{ACM}}, \bibinfo{pages}{582--590}.
\newblock


\end{thebibliography}

\end{document}